\documentclass[preprint2,natbib]{aastex}

\usepackage{graphicx}
\usepackage{amssymb}
\usepackage{amsmath}
\usepackage{enumerate}
\usepackage{epstopdf}
\usepackage{times}

\newcommand{\gr}{$\gamma$-ray}
\newcommand{\kms}{km~s$^{-1}$}
\newcommand{\ie}{i.e.~\/}
\newcommand{\eg}{e.g.,}
\newcommand{\ncl}{$n_{\rm cl}$}
\newcommand{\rcl}{$r_{\rm cl}$}
\newcommand{\dcl}{$d_{\rm cl}$}
\newcommand{\dphot}{$d_{\rm phot}$}

\newcommand{\MP}{$P_{\rm max}$}

\newcommand{\PEW}{polarized equivalent width}

\newcommand{\QEW}{Q-vector equivalent width}
\newcommand{\UEW}{U-vector equivalent width}
\newcommand{\SEW}{Stokes equivalent width}
\newcommand{\linecent}{8000}

\newcommand{\rvii}{400~\kms}

\newcommand{\rvviii}{1600~\kms}
\newcommand{\rvxvi}{3200~\kms}
\newcommand{\rvxxxii}{6400~\kms}
\newcommand{\rvxl}{8000~\kms}

\newcommand{\Ip}{\ensuremath{I_{\rm p}}}
\newcommand{\Pp}{\ensuremath{P_{\rm p}}}

\begin{document}

\title{Spectropolarimetric Signatures of Clumpy Supernova Ejecta}

\author{K. T. Hole,\altaffilmark{1,3} D. Kasen,\altaffilmark{2} 
K. H. Nordsieck\altaffilmark{1}}

\altaffiltext{1}{Astronomy Department, University of Wisconsin-Madison, Madison, WI 53706.}
\altaffiltext{2}{Department of Physics, University of California, Berkeley
and Lawrence Berkeley National Laboratory, Berkeley, CA, 94720}
\altaffiltext{3}{Physics Department, East Tennessee State University, Johnson City, TN, 37614.}

%

%

%
\begin{abstract}
%

Polarization has been detected at early times for all types of
supernova, indicating that such systems result from or quickly
develop some form of asymmetry. In addition, the detection of strong
line polarization in supernovae is suggestive of chemical
inhomogeneities (``clumps") in the layers above the photosphere, which
may reflect hydrodynamical instabilities during the explosion. We have
developed a fast, flexible, approximate semi-analytic code for
modeling polarized line radiative transfer within 3-D inhomogeneous
rapidly-expanding atmospheres. Given a range of model parameters, the
code generates random sets of clumps in the expanding ejecta and
calculates the emergent line profile and Stokes parameters for each
configuration. The ensemble of these configurations represents both
the effects of various host geometries and of different viewing
angles. We present results for the first part of our survey of model
geometries, specifically the effects of the number and size of clumps
(and the related effect of filling factor) on the emergent spectrum
and Stokes parameters.  Our simulations show that random clumpiness
can produce line polarization in the range observed in SNe Ia ($\sim$1-2\%), as well
as the Q-U loops that are frequently seen in all SNe. We have also
developed a method to connect the results of our simulations to robust
observational parameters such as maximum polarization and \PEW~in the
line. Our models, in connection with spectropolarimetric observations,
can constrain the 3-D structure of supernova ejecta and offer
important insight into the SN explosion physics and the nature of
their progenitor systems.

\end{abstract}


%
\section{Introduction}
\label{s:introduction}

Our understanding of supernovae (SNe) has an impact on a wide variety
of astronomical fields, from star formation, interstellar medium
dynamics and chemical evolution of galaxies through cosmological
models and the distance ladder. Indeed, our dependence on the
uniformity of Type Ia SNe (SNe Ia) requires a greater precision of
theory than is generally necessary in astronomy. Yet there remain a
number of unanswered questions in the field that may impact the
accuracy of results obtained from our current models of supernovae.
One important aspect of SNe not yet well understood is their
asymmetry, which may impact their observed brightness as well as the
deposition of kinetic energy and enriched material into their
environment.

We have broad indications of SN asymmetry both from theory and from
observation. High proper-motion pulsars from SNe, \gr\ bursts and
asymmetric remnants support the idea observationally for core-collapse
SNe \citep{lyne94,woosley06,fesen01,hwang04}; the leading theories of
SNe Ia progenitors imply breaking of symmetry due to binarity \citep{wang08}, as do
the hydrodynamical instabilities and and off-center ignition conditions seen in explosion simulations
\citep{khokhlov99,plewa04,roepke07}.  Further, basic expectations
regarding rotation and magnetic fields suggest asymmetries
in the progenitor star structure with potentially complicated effects on the
resulting ejecta.

Observational probes of SN asymmetries have been challenging. All SNe
yet seen in the modern era of astronomical detection have not been
resolvable for years after the explosion -- the first resolved images
of SN 1987A, the closest in modern times, with the Hubble Space
Telescope came in 1991 \citep{podsiad91}.  Further, the most common
diagnostic of asymmetry in unresolved sources -- line profiles -- are
not practical in SNe, where ejecta velocities are typically on order
of 10,000-20,000 \kms\ 
or more. Therefore the most
effective way of probing the asymmetries in SNe is with
spectropolarimetry.

Polarization tells us about the asymmetries of an unresolved
object. In SNe, polarization is generally thought to arise from
electron scattering in the last few optical depths within the
photosphere. If the source is completely symmetric, all polarization
angles will be present in equal quantities, and we will measure no net
polarization -- the equivalent of ``natural light." If, on the other
hand, the source photosphere is asymmetric, or if some of the
polarized light is blocked by an asymmetric opacity above the
photosphere, different polarization angles will be present in
different strengths, and we will detect a net polarization
\citet{kasen03}.

\emph{Spectro}polarimetry has the potential to
give us even more insight into the structure of SN. 
In supernovae, the velocity of individual mass elements are set within the first few days after the initial explosion. The mass of the ejecta quickly stratifies into a pseudo-Hubble flow, where distance from the center of the explosion is proportional to velocity. Thus, in a SN spectrum, observations at a given wavelength will correspond to the same mass-elements over time.
By measuring the
variation of polarization with wavelength, we can observe asymmetries
as a function of radius, as well as the effects of depolarization by line scattering, giving us
information about chemical and density inhomogeneities within the
ejecta.
 
Many observers have contributed to our knowledge of the
spectropolarimetric characteristics of SNe
\citep[\eg][]{leonard06,maund07,chornock08,tanaka09}.  The results are
well discussed in \citet{wang08}, so here we will only state the most
general conclusions: Core-collapse supernovae show evidence of an
underlying axisymmetric explosion mechanism, and also show significant
deviations from that axisymmetry in individual cases. SNe Ia show
little continuum polarization and thus are likely close to globally
symmetric. SNe Ia absorption lines, on the other hand, are often
associated with strong polarization peaks. This can be interpreted as
being due to clumpy structures in the outer layers, composed of
intermediate mass elements, which asymmetrically block the
photosphere.

To understand the host geometry that produced a given
spectropolarimetric signature, it is necessary to model radiative
transfer through possible ejecta structures and match the results to
observations.  The problem of modeling SN spectropolarimetry has been
approached in a variety of ways over the past few decades. 
Most codes for modeling spectropolarimetric radiative transfer within the ejecta of
SNe Ia, such as HYDRA \citep{hoeflich05}, and SEDONA \citep{kasen06} use a Monte Carlo approach to line and  continuum radiative transfer and \gr\  deposition within aspherical ejecta to calculate time-dependent spectropolarimetry from a variety of viewing angles.
Another approach to SN modeling, developed by Jennifer
Hoffman, specifically addresses the interaction of core-collapse SNe
with circumstellar material in Type IIn SNe \citep{hoffman05}.

These codes are highly detailed and powerful. Yet this complexity also
means that they are computationally intensive, making large-scale
parameter studies impractical. Since polarization is a second order
effect of the geometry, different source configurations may produce
the same polarization signature. It is therefore necessary to
understand the statistical likelihood of different source
configurations producing a given observational signal. This requires
systematic modeling of the expected signal given a wide range of
source geometry. The large number of model configurations required for
reasonable statistics, the three-dimensional nature of the problem,
and the wavelength dependent polarization effects combine to pose a
significant computational challenge.

Since our goal is to perform just such a parameter study, we have
taken a different approach to supernova spectropolarimetric radiative
transfer modeling. In the interest of simplicity, we have chosen to
focus our simulations on one potential source of SN asymmetry: regions
of enhanced line opacity (``clumps") in the layers above the SN
photosphere suggested by the detection of strong line polarization in
SNe.  If present, the nature of this clumpiness has implications for
explosion models of SNe.

The goals of this paper are therefore to calculate the \emph{line}
polarization from inhomogeneous distributions of elements in the
ejecta, and to statistically examine how it relates to clump
properties.  We therefore systematically study the effects of various
numbers and sizes of clumps, placed randomly to recapitulate the
effects of a range of explosion scenarios as well as those of
line-of-sight orientation. To do this, we have developed a fast,
flexible approximate semi-analytic code for modeling radiative
transfer within 3-D inhomogeneous ejecta. In this paper, we present
this code and our methods for statistically analyzing large numbers of
simulated spectra, and use them to explore a variety of possible
clumpy ejecta scenarios. From the ensemble of resultant Stokes spectra
we can determine what ranges of parameters predict observable
characteristics that can constrain aspects of the host geometry.

In \S \ref{s:code}, we describe our code and how we extract observable
trends from an ensemble of calculated spectra. In \S 3, we present the
results of our analysis of the effect of number of clumps and clump
size. In \S 4 we summarize the current work and discuss the code's
future potential.

\section{Radiative Transfer Code and Analysis Methods}
\label{s:code}
%

\subsection{Model Geometry and Assumptions}
\label{ss:geometry}

The most basic simplifying assumption in the modeling of supernova
ejecta is that all the material is moving purely radially and with the
velocity of the material increasing linearly with distance from the
center of the explosion.  We can make this assumption because the
energy that feeds the SN expansion is extremely large, and most is
input almost instantaneously (on order of a few seconds). Initial
shocks and entrained radioactive material will affect the structure of
the ejecta for a few days, but such effects will be minimal by the
time the SN is detected,  on order of 10 days after the explosion  \citep{arnett96}.
Further, the expanding ejecta is unlikely to encounter any
significant circumstellar material at this point in expansion, as that
material appears to be cleared out by radiation and other
pre-explosion dynamics in the years or centuries before the
explosion. The main exception to this may be SN IIn \citep[and
references therein]{hoffman08}, which we do not address in our models.
The motion of the ejecta can therefore be considered ballistic.

\begin{figure}[htbp]
\begin{center}
\plotone{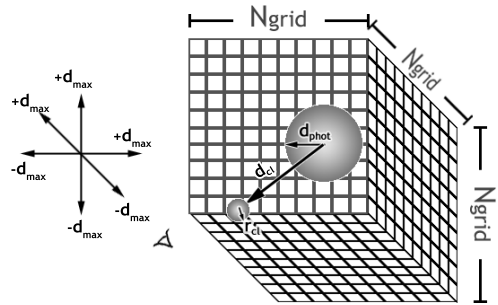}
\caption{Model grid geometry.}
\label{f:grid}
\end{center}
\end{figure}

Our model therefore begins with an $N^{3}$ grid extending from
$-d_{\rm max}$ to $+d_{\rm max}$ in each dimension in velocity space
(see Fig. \ref{f:grid}). By the time a SN becomes observable, the
ejecta will have stratified by velocity so that this grid is
equivalent to a spatial grid at a given time. Because of the linear
dependence of velocity on distance, this structure is sometimes
referred to as a pseudo-Hubble flow, and has many of the same
characteristics of Hubble expansion. Our supernova is centered at
$d=0$ and has a photosphere at \dphot, outside of which is the ejecta
through which our radiative transfer is calculated.  
The SN ejecta is first initialized to a symmetric,
smooth optical depth that decreases as a power law from the
photosphere ($\rho=\rho_{0}r^{-n_{ej}}$).  (See Table \ref{t:model}
for common parameter values used in all simulations reported in this
paper.) Regions of enhanced opacity (clumps) are then added, with
their particular parameters selected randomly from a range set for a
given simulation.  Each simulation consists of 1000 spectra in an
``ensemble" of realizations with the same parameter ranges. Clump
parameters that can be randomly assigned in this way include the
number, central optical depth, radii and distance from the photosphere
of the clumps (\ncl, $\tau_{\rm cl}$, \rcl~and \dcl, respectively).
We do not specifically vary line-of-sight in our results, because
given a large ensemble, the random placement of clumps will
recapitulate a variety of lines of sight. This simplification greatly
reduces computational and analytical complexity.

Our Stokes spectra have $N$ wavelength bins centered around a central
wavelength, in this case \linecent~\AA. Note that the exact wavelength
is only relevant to the background blackbody spectrum; for these
simplified models the ``line'' is pure trapped resonance scattering
and particular line physics are not addressed. $N$ here is the same as
the grid dimension.

Our radiative transfer is done in two parts.  We begin with the intensity and polarization of the continuum flux, which for a spherically symmetric photosphere can be represented in polar coordinates as 
\begin{equation}
\label{eq:dk1}
\left(
\begin{array}{ccc}
I(r,\phi) \\ 
\hat{Q}(r,\phi) \\
\hat{U}(r,\phi) \\
\end{array}
\right)
= 
\left(
\begin{array}{ccc}
\Ip(r) \\ 
\Ip(r) P_p(r) \cos 2 \phi \\
\Ip(r) P_p(r) \sin 2 \phi \\
\end{array}
\right)
\end{equation}
where $\Ip(r)$ is the specific intensity and  $\Pp(r)$ the percent polarization emerging from an impact parameter $r$. The Stokes parameter corresponding to circular polarization, $\hat{V(r,\phi)}$, is taken to be zero, and thus not included in our equations. This is consistent with observations,
as no circular polarization has yet been detected in a supernova \citep{wang08}. 
We also use use the Stokes parameter notation given in
that paper, where $I$, $\hat{Q}$ and $\hat{U}$ 
represent the measured vector components, while $Q=\hat{Q}/I$ and
$U=\hat{U}/I$ have been normalized. 
Integrating the intensity over a plane perpendicular to the line of sight gives the Stokes flux at a specific wavelength
\begin{equation}
\label{eq:dk2}
\begin{split}
I &=\int_0^{2 \pi} d \phi \int_{d_{min}}^{d_{max}} r dr~ e^{-\tau(r,\phi)} I_p(r) 
 + ( 1 -  e^{-\tau(r,\phi)}) S
\\
\hat{Q} &=
\int_0^{2 \pi} d \phi \int_{d_{min}}^{d_{max}} r dr~ e^{-\tau(r,\phi)} I_p(r) P_p(r) \cos 2\phi \\
\hat{U} &=
\int_0^{2 \pi} d \phi \int_{d_{min}}^{d_{max}} r dr~ e^{-\tau(r,\phi)} I_p(r) P_p(r) \sin 2\phi  
\end{split}
 \end{equation}
where $r_p$ is the photospheric 
radius,  $d_{min}$ is the impact parameter at
which the given plane intersects the photosphere (or zero if the photosphere is not
intersected)
and $S$ is the line source function, assumed to be pure scattering, $S = W(r,\phi) I_p$ where $W$ is the geometrical dilution factor. The plane located at distance $z$ along the line of sight (with a  projected velocity $v_z$) corresponds to a wavelength $\lambda = \lambda_0 ( 1 + v_z/c)$. In theory, each of these function could have wavelength dependence, but 
for simplicity over such a small wavelength range we do not vary any of these functions except by letting $\Ip(\lambda)$ vary according to a blackbody spectrum.

In our simulations, $\Ip(r)$ and $\Pp(r)$ are set by the results of a spherical Monte
Carlo calculation. Note that for a spherical photosphere, the
polarization at the limb is higher than in the plane parallel
calculation of \citet{chandra60},
used before in e.g., \citet{shapiro82}.
The values of the Q and U Stokes vectors across the observer-oriented
face of the photosphere are taken from the model calculated in
\citet{kasen03}. 

Equation \ref{eq:dk2} represents the effect of line opacity above the photosphere in obscuring or
diluting continuum flux is calculated. Operationally, for the radiative transfer
through the ejecta, we calculate the interacting wavelength for each
cell in the 3-D velocity grid using the Sobolev approximation
\citep{sobolev60} for line opacity. The light from the photosphere
that passes through that cell is then attenuated or scattered at the
resonant frequency by the amount determined by the cell optical depth.
The combined effects of each cell on all wavelengths gives us our line
profile, and asymmetric depolarization of the photospheric flux will
give us a spectropolarimetric signal.

Note that we make two main assumptions in this model: First, that
there is a clear division between the electron scattering photosphere
and the line forming region of clumps.  In reality, there is no sharp
division, and some electron scattering will be coincident with the
line scattering. However, if the clumps can be represented as
sufficiently detached from the photosphere, this model will still
provide relevant results \citep[see discussion in ][]{kasen03}.

Second, we make the common assumption that all \emph{line} scattered
light is depolarized.  Though this is certainly a simplification of
the physical picture, it is a reasonable approximation in most
cases. The reasoning given in \citet{hoeflich96} is that collisional
timescales are less than those for absorption and re-emission in the
ejecta; thus the random scattering processes will erase incoming
geometrical information for an absorbed photon. We add that even if
collisions are not dominant, line polarization is unlikely to be
significant because lines are at best weakly polarizing, comparatively \citep{hamilton47}
with the greatest polarization in the side scattering case. In our
geometry, photons that are {\em side} scattered into the line of sight
will be seen in the emission portion of the line, while observed
polarization is in the absorbed part of the line. Further, any light
line-scattered into the line of sight will be a small contribution to
the total flux, making such a signal harder to detect.

Thus in our model, polarized rays of light from a pre-computed
spherical photosphere are obscured by clumps of depolarizing line
opacity.  Because the clumps block the photosphere in an asymmetrical
way at different velocities, they give rise to polarization over the
line flux absorption feature.


\begin{table}
\begin{center}
\caption{Model parameters used for simulations in this paper. Sizes are given in velocities, which are proportional to distances in the homologous flow of SN ejecta (see \ref{s:introduction}).\label{t:model}
\vspace{.1cm}}
\begin{tabular}{cc}

\tableline
Parameter & Value \\

\tableline\tableline
N & 100 \\
$d_{\rm max}$ & 20,000 \kms \\
\dphot & 10,000 \kms \\
T$_{\rm BB}$ & 13,000 K \\
$\tau_{\rm cl}$ & 10 \\
$\lambda_{\rm cent}$ & \linecent \AA \\
$n_{\rm ej}$ & 8 \\

\tableline
\end{tabular}

\end{center}
\end{table}

\begin{figure}[htbp]
\begin{center}
\includegraphics[width=3in]{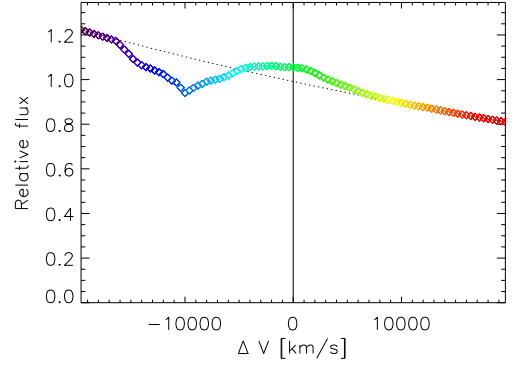}
\includegraphics[width=3in]{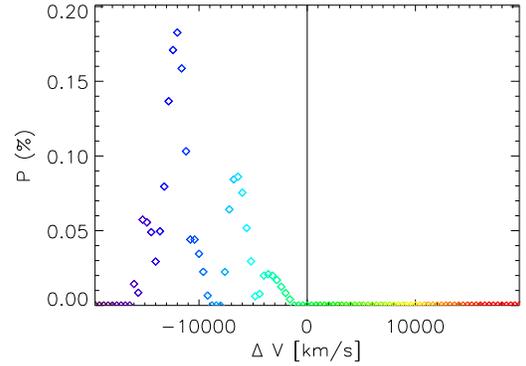}
\includegraphics[width=3in]{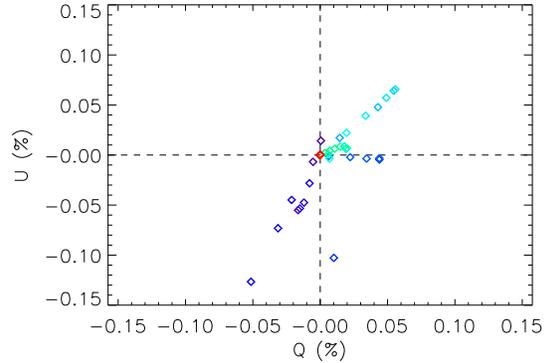}
\caption{A calculated line profile, spectropolarization profile and
Q-U diagram for the \ncl=32, \rcl=\rvviii~case. This particular
realization has several clumps in the line-of-sight and therefore has
a number of broad polarization peaks and distinct variations from a
single dominant axis in the Q-U plane.}
\label{f:spec_lots}
\end{center}
\end{figure}
\begin{figure}[htbp]
\begin{center}
\includegraphics[width=3in]{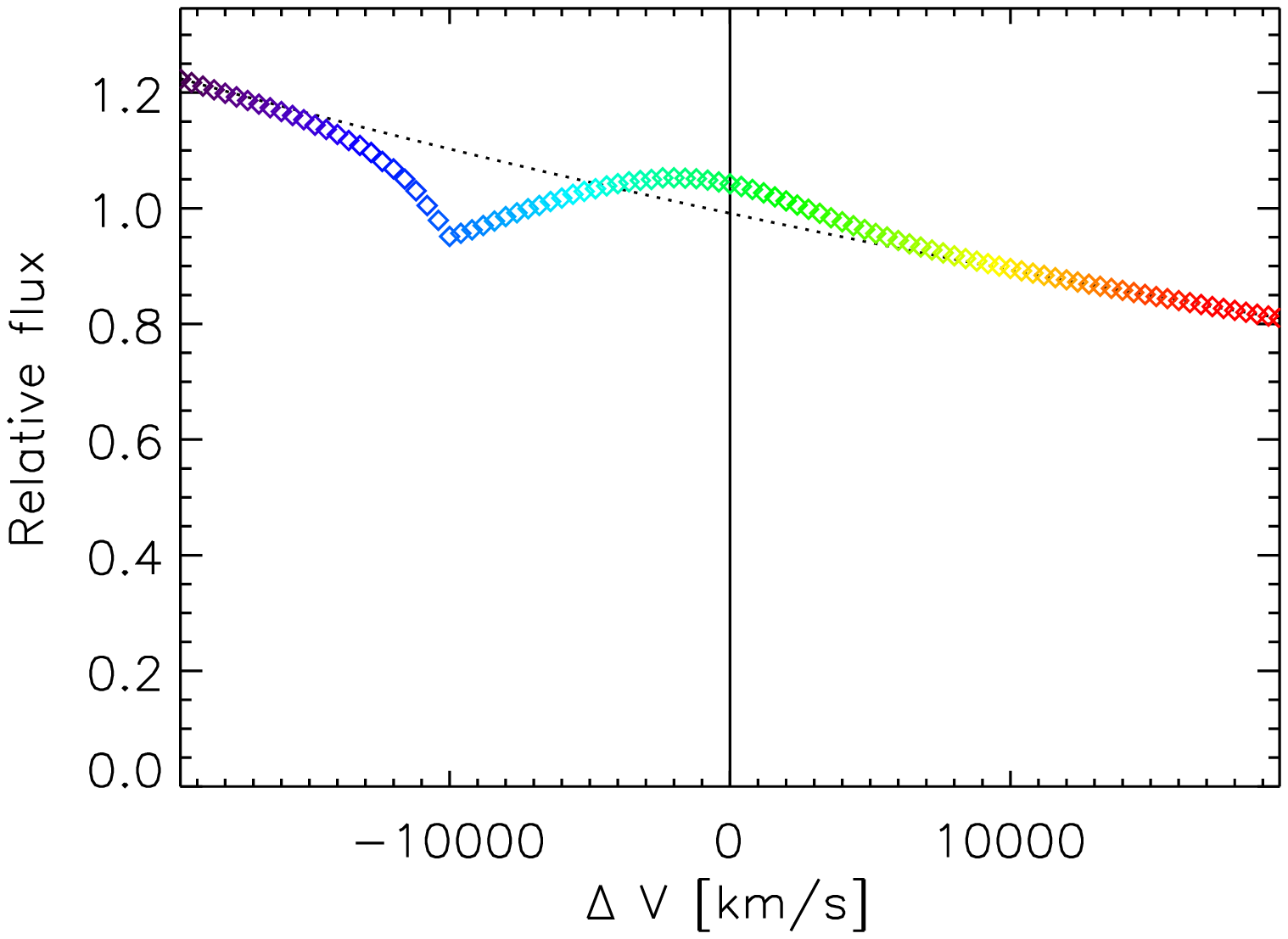}
\includegraphics[width=3in]{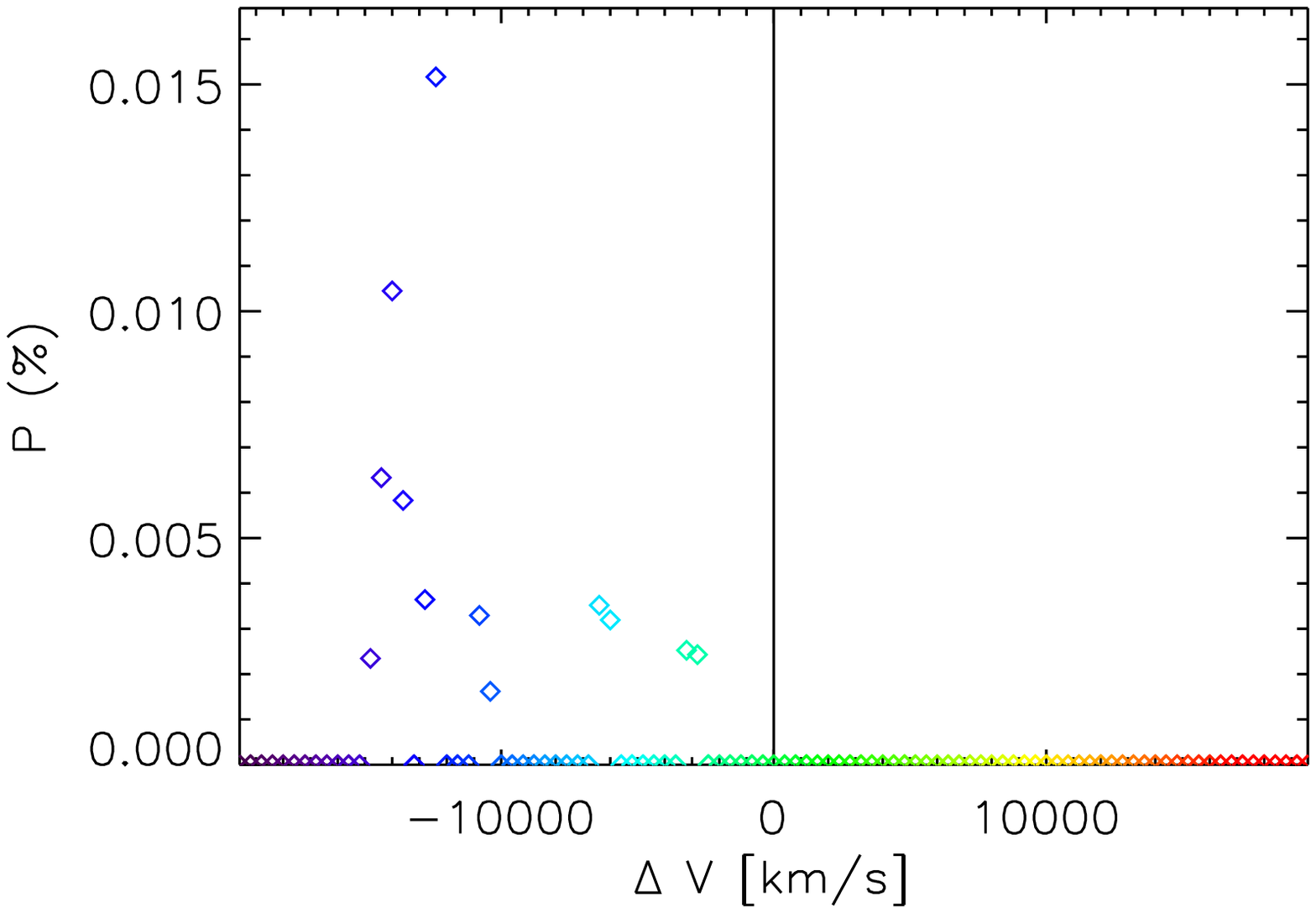}
\includegraphics[width=3in]{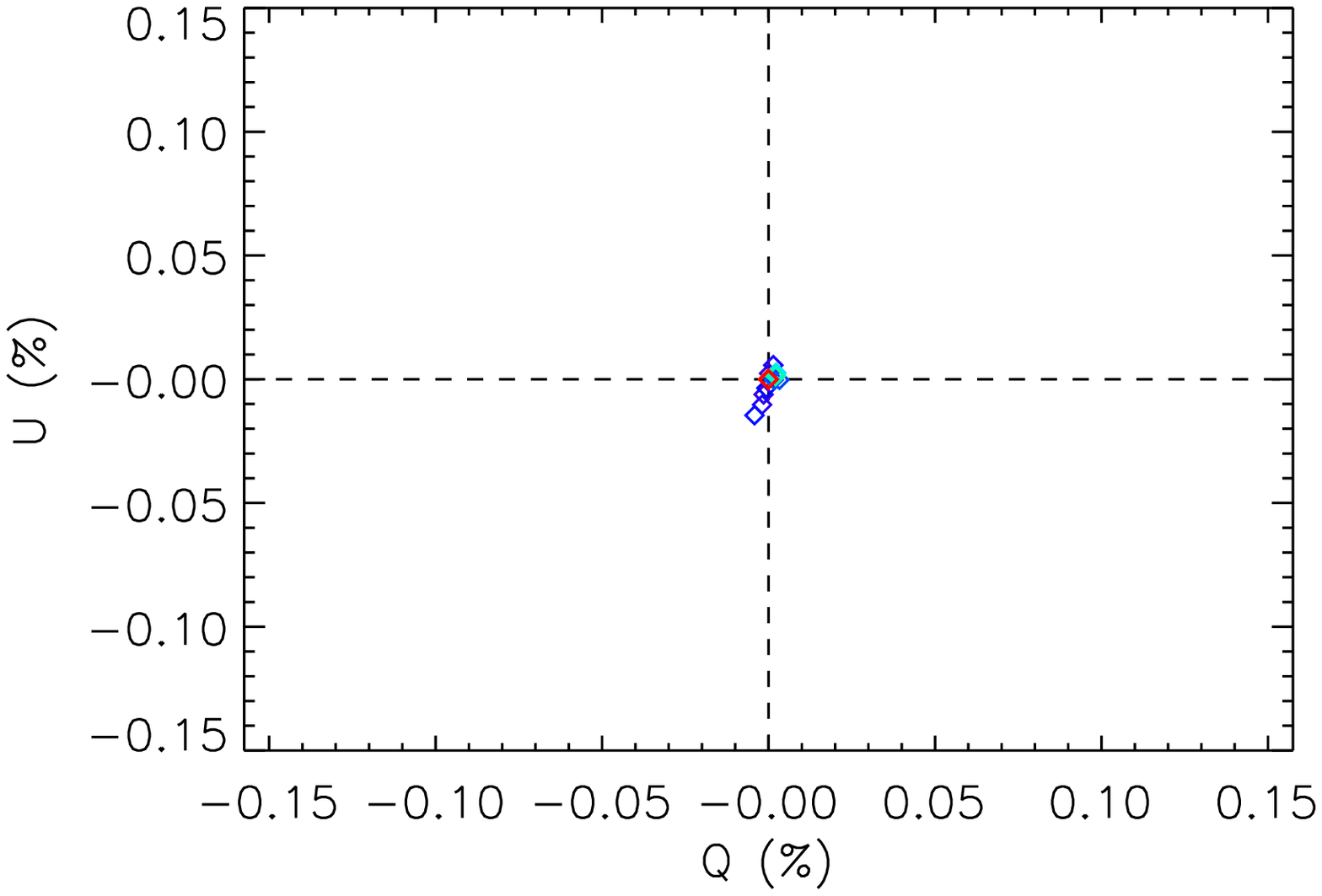}
\caption{A calculated line profile, spectropolarization profile and
Q-U diagram for the same clump configuration as in figure
\ref{f:spec_lots} but with \rcl=\rvii~instead of \rvviii. Note that
the polarization peaks are significantly narrower and $\gtrsim10$
times smaller in magnitude.}
\label{f:smaller_lots}
\end{center}
\end{figure}
\begin{figure}[htbp]
\begin{center}
\includegraphics[width=3in]{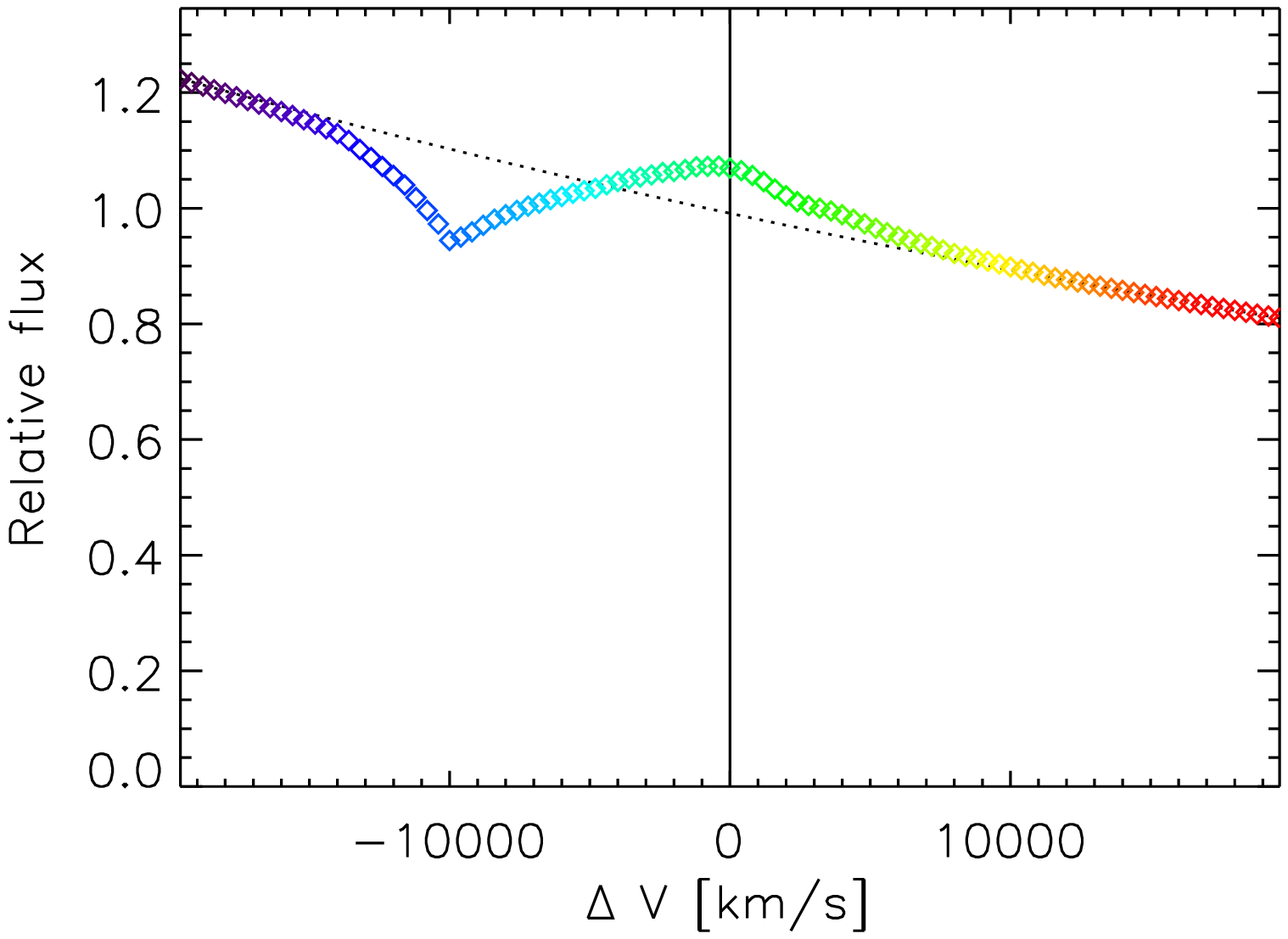}
\includegraphics[width=3in]{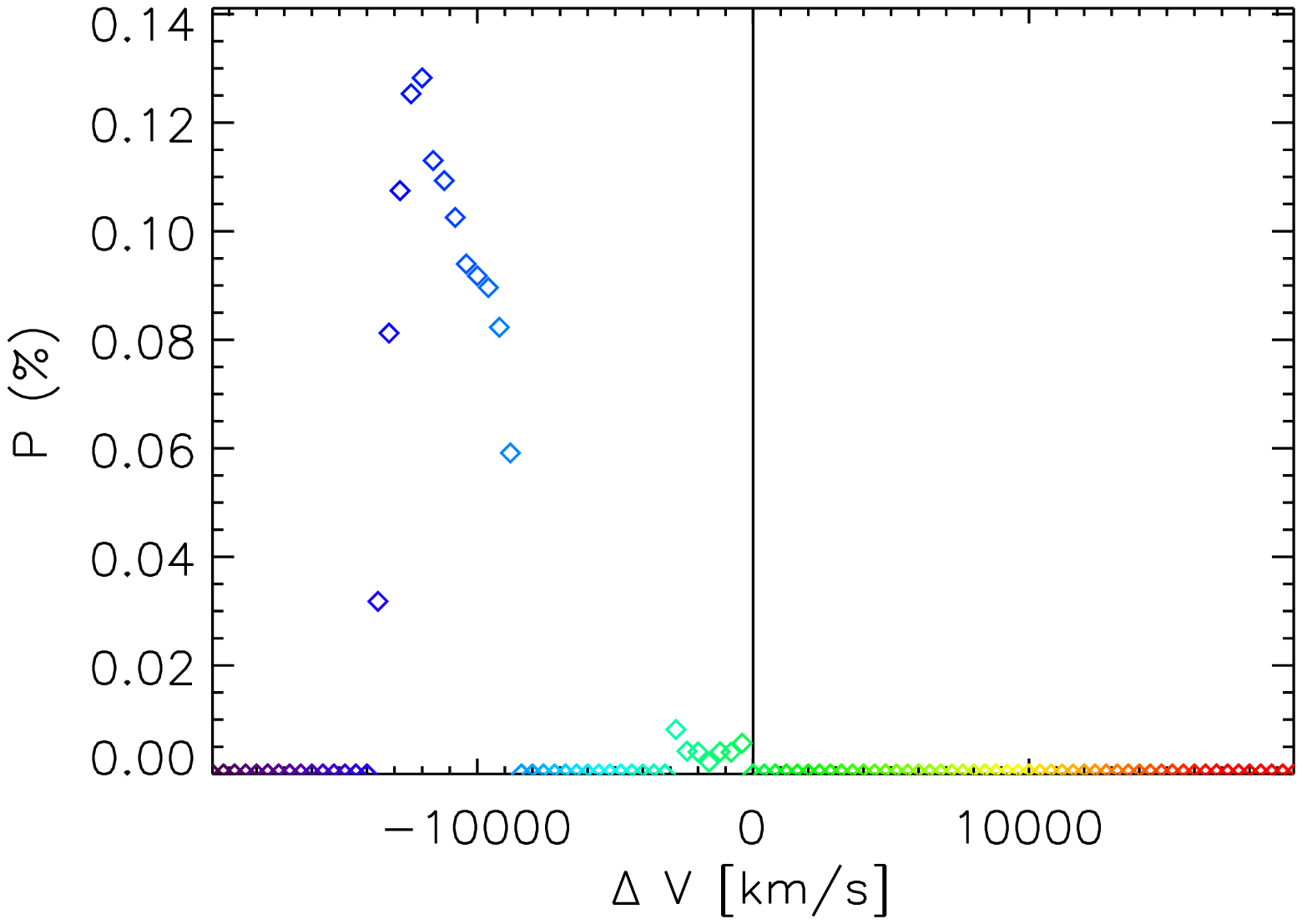}
\includegraphics[width=3in]{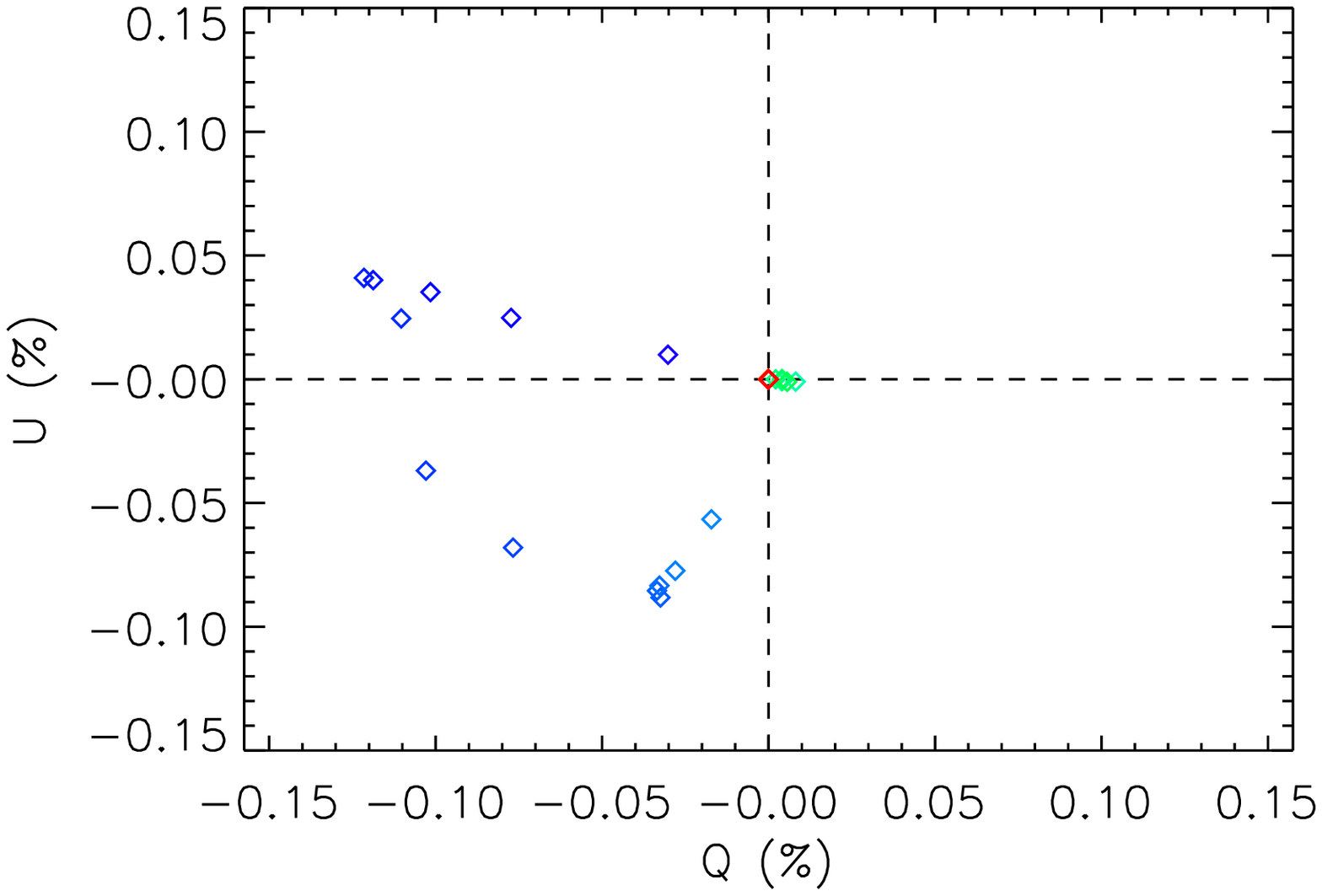}
\caption{Another calculated line profile, spectropolarization profile
and Q-U diagram for the \ncl=32, \rcl=\rvviii~case (\eg~it is from the
same ensemble as the example in figure \ref{f:spec_lots}). This
particular realization produces a classic ``loop'' in the Q-U plane. The difference that leads to this loop is hinted at by the plots of percent polarization: in figure \ref{f:spec_lots}, the clumps are separated in velocity space, so the Q-U diagram shows distinct deviations from zero, with a separate position angle for each clump. In contrast, the clumps in this realization are overlapping in velocity and wavelength, so the Q-U diagram shows a smooth varying from one clump's position angle to another, creating a ``loop" structure.}
\label{f:spec_loop}
\end{center}
\end{figure}

Results from our code have been compared to those of Kasen's previous
semi-analytic codes \citep{kasen03} with the same host configurations,
to good agreement. See \citet{hole09} for more details. Flux and
polarization spectra and Q-U diagrams for three sample realizations
are shown in Figures \ref{f:spec_lots}, \ref{f:smaller_lots} and
\ref{f:spec_loop}.

\subsubsection{Regimes in \ncl-\rcl\ Parameter Space}
\label{ss:limits}

The code can be run with any number or size of clumps, but it is worth
examining what is physically implied by different regions of
\ncl-\rcl\ parameter space. In the small-clump limit, we are modeling
small regions of enhanced concentration of an element, surrounded by
smooth ejecta which contribute the bulk of the line profile but no net
polarization. As the clumps become larger or more numerous, the model
more closely resembles large scale variations in the chemical
composition of the overall ejecta structure. Eventually, however, the
simulations with the most and largest clumps will completely fill the
model grid, and represent a substantial change to the nature of the
ejecta, rather than a perturbation of it, to a degree not supported by
observation.

\citet{thomas02} attempted to constrain clump size and filling factor
using line flux profiles only. Comparing results of a numerical clump
model to observations of the Si II line, they concluded that clumps
larger than 0.08 times the photospheric radius (\rvviii\ in our model)
would create line-of-sight variations larger than what was seen. Their
model was substantively different from ours, however, as they
represented clumps as optically thin and surrounded by zero line
opacity. The line profile is thus due entirely to the clumps, rather
than clumps being a perturbation on the line structure. Therefore in
our model, the line-of-sight effects will be smaller even for large
clumps, and should have less variation with angle as long as there are
enough clumps to fill a majority of the ejecta.

In an attempt to quantify the useful parameter limits in our model, we
use a metric based on the photodisk covering factor (PCF) used by
\citet{thomas02}, which they define as the ratio of the area obscured
by clumps contained in the projected photodisk area to the total
photodisk area for each plane of common velocity along the
line-of-sight. The sum of the ratios over all velocities gives a
measure of volume filling factor for the ejecta. For these
simulations, there are 50 velocity planes on the observer side of the
grid, so the sum of the ratios has a maximum of 50 if each plane has
100\% of the available photodisk covered.

\begin{figure}[htbp]
\begin{center}
\plotone{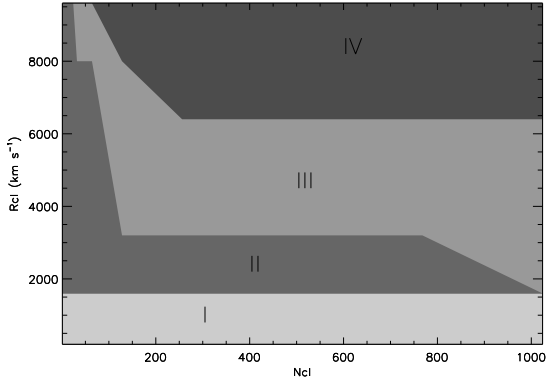}
\caption{Regions of \rcl-\ncl\ parameter space. Region I represents
the small-clump limit, with \rcl~$\leq$ \rvviii.  Regions II and III
represent medium-sized clumps, with a smaller ejecta filling factor in
Region II and greater clump overlap in Region III. Region IV is the
filled-ejecta, theoretical limit of the model. }
\label{f:regions}
\end{center}
\end{figure}

We therefore delineate four regions of parameter space in our model, as shown in Fig. \ref{f:regions}:
\begin{enumerate}[I]
\item The small-clump regime, \rcl\ $\leq$ \rvviii, used by
\citet{thomas02}
\item Medium-clump regime with smaller filling factor (PFC $<$ 38) 
\item Medium-clump regime with larger filling factor (38 $\leq$PCF $<$ 48) 
\item Full ejecta (PFC $\geq$ 48)
\end{enumerate}

Region I has small clumps and is numerically and physically
plausible. Regions II and III have more substantial clumpiness. In
Region II the ejecta is less full, which might lead to greater
line-of-sight variation than is seen in observations. Region III has a
higher filling factor and thus fewer line-of-sight effects, but also
results in more clump overlap, implying the possibility of more
complex polarization effects that may not be completely captured with
our simplified radiative transfer. Region IV represents a full
numerical grid and is not likely to occur in SNe but is shown here to
demonstrate the behavior of the model in extreme cases.

\subsection{Analysis of Spectra}
\label{s:anal_spectra}
%

Given that each simulation creates an ensemble of 1000 individual
Stokes spectra, we developed tools to analyze and visualize the
results in a statistical way. We therefore calculate certain salient,
observable numbers from each spectrum, and determine the ``expected
value'' for each configuration. The numbers calculated from each
spectrum are minimum flux and the velocity and polarization where it
occurs; maximum percent polarization ($P$, see eqn.~\ref{eq:p}) at any
frequency in the spectrum (\MP); \PEW, \QEW, \UEW, and \SEW~(PEW, QEW,
UEW and SEW, see eqn.~\ref{eq:EWs}); and theoretical covering fraction
(TCF, eqn.~\ref{eq:tcf}).

The polarization is defined as
\begin{equation} \label{eq:p}
P =  \frac{\sqrt{{\hat{Q}}^2+{\hat{U}}^2}}{I} = \sqrt{Q^2+U^2}
\end{equation}
We also define a polarized version of line equivalent width: 
\begin{equation} \label{eq:EWs}
\mathrm{[X]EW} =  
\sum \% \mathrm{X}_{\lambda} 
\left( \frac{\mathrm{I}_{c,\lambda}-\mathrm{I}_{\lambda}}
 {\mathrm{I}_{c,\lambda}} \right) \Delta \lambda, 
\end{equation}
where [X] can be P, Q, or U and $\mathrm{I}_{c}$ is the continuum
flux.  Since QEW and UEW should each be sampling the same underlying
distribution even if their values differ in a given spectrum, we add
these separate distributions together to double the sample size. We
call this combined value ``Stokes equivalent width,'' or SEW.  In this
paper, we focus primarily on the analysis of \MP~and PEW results.

The concept of the ``theoretical'' covering fraction is also worthy of
more discussion. It is useful to have a way of calculating a covering
fraction in order to facilitate comparison to the results of explosion
codes; however the correct measure of covering fraction in our
simulations is not as simple as one might expect. First, our randomly
placed clumps may not be in the line of sight, and therefore may not
in that sense be ``covering'' the photosphere from the observer's
point of view. However, to most easily compare our results with the
predictions of explosion codes, we need to account for the total
number of clumps. Further, because our clumps are distributed in
velocity space, the covering fraction will be different at different
frequencies depending on where the Sobolev line interaction
occurs. The covering fraction at each wavelength in the line of sight
can be calculated (\eg the ``photodisk covering fraction'' of section
\ref{ss:limits}) however this measure is not ideal here because it is
a) a spectrum rather than a single number, b) subject to line-of-sight
effects and c) more complicated to compare to explosion models. We
therefore have chosen to use the theoretical covering fraction, which
we define as follows:
\begin{equation} \label{eq:tcf}
\mathrm{TCF} =  \sum_{N_{\rm cl}} \frac{\pi r_{\rm cl}^2}{4 \pi d_{\rm cl}^2}
\end{equation}

In words, we find the fraction each clump subtends of the total area
of the sphere located at the distance of the clump, and then take the
sum of these fractions for all clumps. There are limitations to this
method, particularly in that our clumps may overlap either within the
opacity grid or in line of sight, and may therefore overestimate the
covering fraction, potentially creating a TCF that is greater than
one, even if there are some locations or wavelengths where the
photosphere is unobscured. It does, however, create a single,
conceptually clear number to use in calculations and comparisons.

\subsection{Analysis of Ensembles}
\label{s:anal_ensem}
%

Once the appropriate metrics are calculated for each individual
spectrum in an ensemble, we can perform statistical analysis on the
results. If we have 1,000 realizations of a given set of parameters,
we have 1,000 predictions of, for instance, the maximum polarization
that could be produced by that configuration. A Gaussian fit to a
histogram of the maximum polarizations thus produced allows us to
predict the likelihood of a given polarization being produced by that
geometry. We can also use such model parameters as number of clumps
(\ncl), clump size (\rcl) or TCF to connect these geometries to those
predicted by different SNe explosion modeling codes, and thereby
determine what spectropolarimetric signatures these explosion codes
would imply in our simulations.

\begin{figure}[htbp]
\begin{center}
\includegraphics[width=3in]{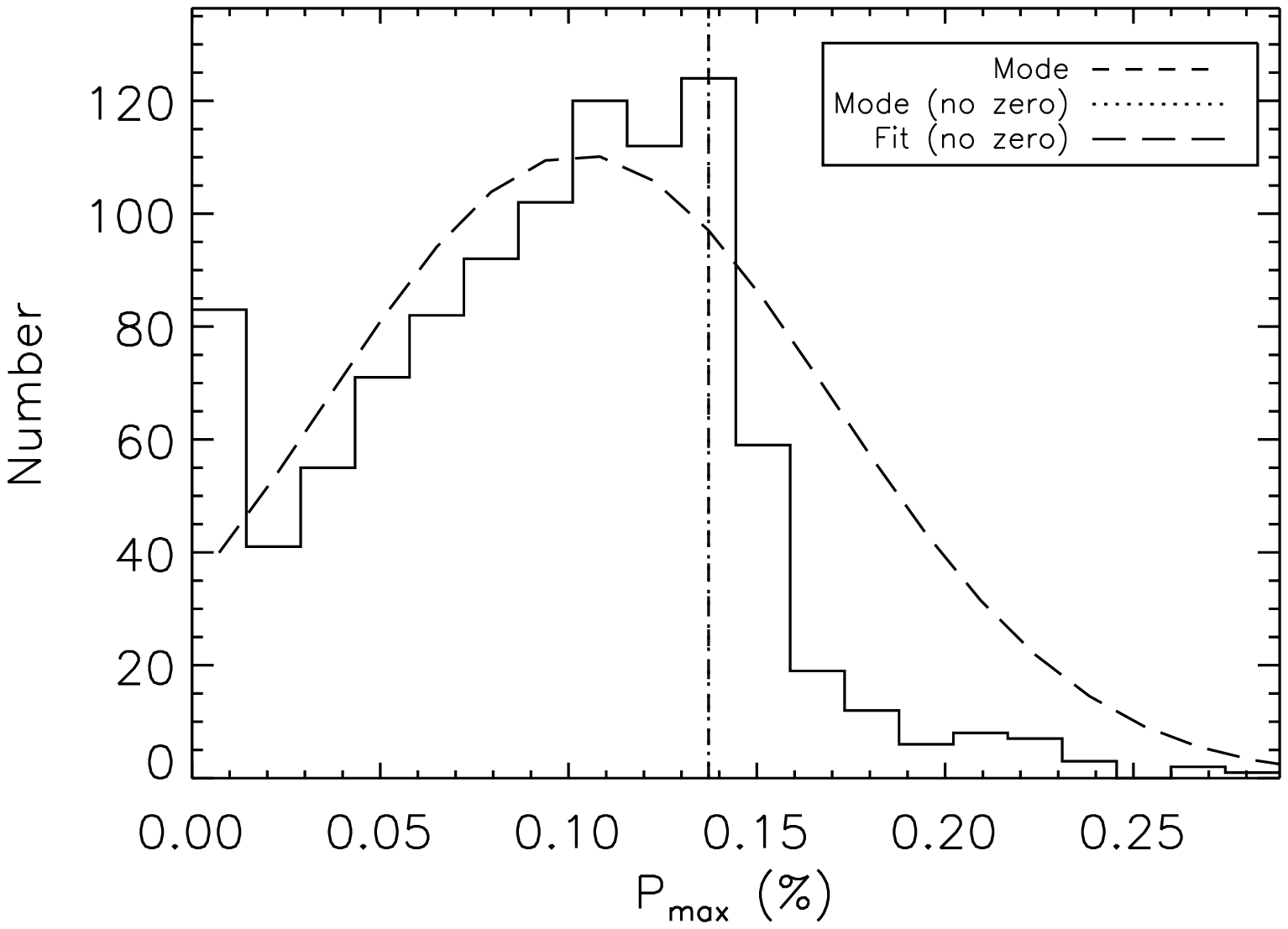}
\includegraphics[width=3in]{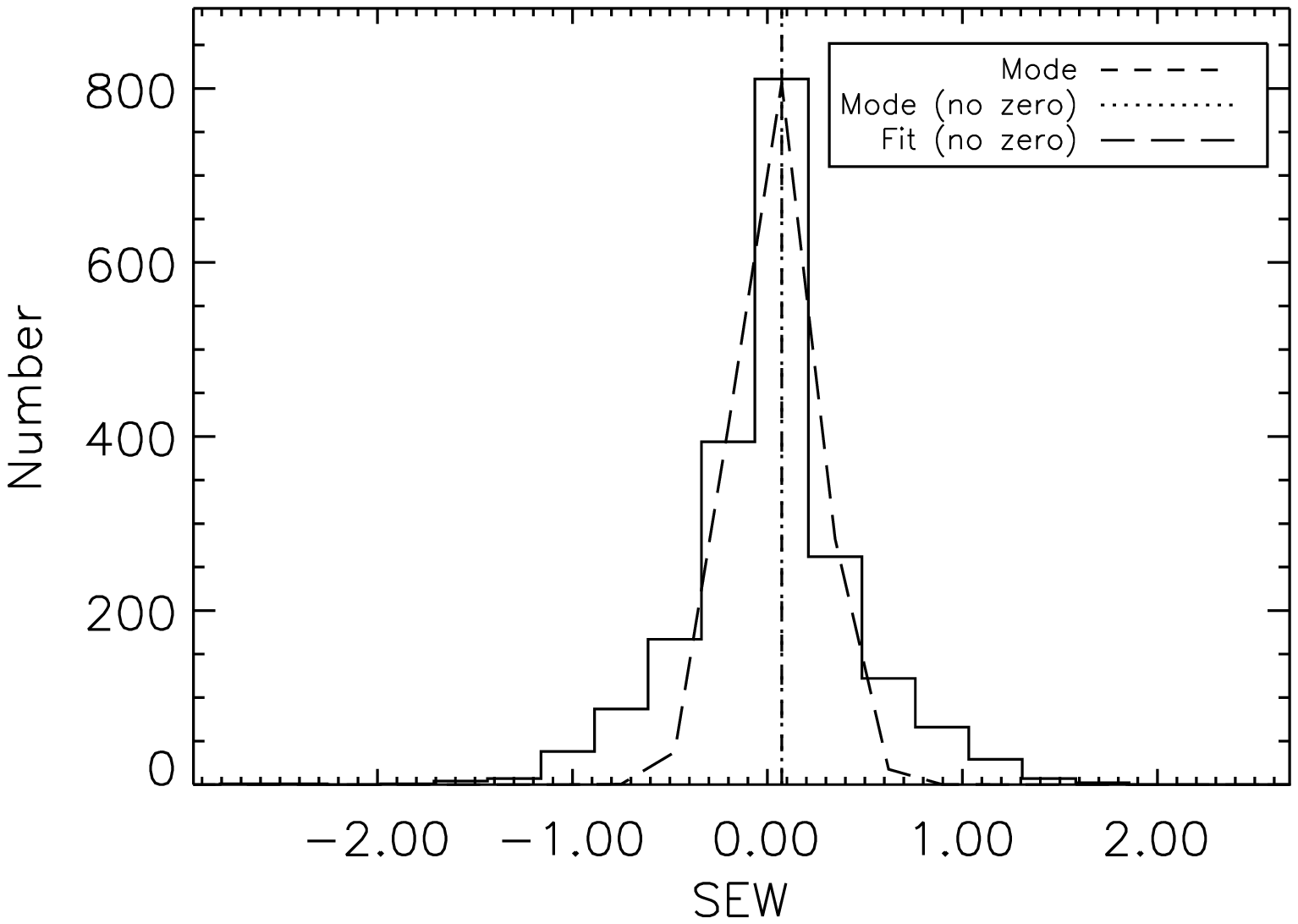}
\caption{The histogram and Gaussian fit for the maximum polarization
produced at any point in the line profile (above) and \SEW~(below) for
an ensemble with \ncl=16 and \rcl=\rvviii. In the \MP~plot, the peak
at zero represents the combination of symmetrically placed clumps and
instances where no clumps appear in the line of sight. The Gaussian
fit was set to avoid this peak; the mode of the distribution is
calculated for the full distribution, and again excluding the peak at
zero. In this case, both modes are the same.  Note that the Gaussian
fit is a better representation of \SEW~data than for \MP, as expected,
because \SEW~should be scattered about zero. } 
\label{f:sampleMP_FQU}
\end{center}
\end{figure}

In Figure \ref{f:sampleMP_FQU} we show some of the statistical results
for an ensemble, namely the histograms of \MP~and SEW. In this
ensemble, clumps were randomly distributed between the photosphere and
the edge of the grid, and with \ncl=16 and \rcl=\rvviii. Additional
parameters are as in Table \ref{t:model}.

Analysis of such histograms can lead to insights into the factors
contributing to the overall distribution. In this ensemble, we see
three main contributions. First, the peak at zero polarization
represents a combination of symmetrically placed clumps and instances
where the clumps were randomly placed so that none of them lie in the
line-of-sight. Between zero and $\sim$0.15\% we see a steadily
increasing slope that make up the bulk of the asymmetric population,
following the expected properties of clump distribution.  Third, in
addition to this bulk population, there are a small number of outliers
in the distribution. These outliers have the potential to demonstrate
interesting though unlikely configurations, and possibly to
distinguish between similar clump models.

In addition to analysis of individual ensemble distributions we seek
to understand the way our results change with variations of simulation
parameters. To see trends in metrics between ensembles, we must find
ways to characterize each distribution in ways easily comparable to
others with different parameters. One measure we use is a Gaussian fit
the distribution, which gives a center and a width.  The true mode of
the distribution is also calculated, and finally, for some metrics, we
find the mode of the distribution if zero is excluded. This second
mode can be useful in distributions that can have a peak at zero due
to, for example, no clump falling within the line-of-sight, which is
distinct from the behavior of the metric when any clumps are in
position to affect the observed signal. This second, non-zero mode
would therefore be useful for metrics that are not expected to be zero
when there are any asymmetries present (\eg~\PEW) but not for metrics
where expected values should be distributed about zero (\eg~$Q_{\rm
max}$).

As an example, we again take the case of \MP~in the line as shown in
Figure \ref{f:sampleMP_FQU}. Here, the Gaussian fit is not a complete
representation of the underlying distribution. Indeed, since P is an
intrinsically positive quantity, the distribution of P measurements in
an ensemble should not be Gaussian by definition.  For this metric,
the mode seems to be a better measure of the behavior of the
distribution produced by our simulation. Unfortunately, the mode
cannot capture the width of the distribution, only its most likely
value.

In contrast, there are some metrics where the distribution of
measurements is more usefully represented by the Gaussian fit than by
the mode. Both Q and U individually are expected to be randomly
distributed about zero, and thus the combination of the two
distributions should be as well.  We therefore expect that this data
will be more accurately represented by a Gaussian fit. In Figure
\ref{f:sampleMP_FQU} we show that the distribution of \SEW~for the
same ensemble is indeed more accurately fit by the Gaussian than was
the \MP. Here the width of the fit is a more useful quantity than the
mode, which will be at or very close to zero.

\epsscale{1.} 
\epsscale{1.}

\subsection{Inter-Ensemble Parameter Studies}
\label{ss:parstud}
%

Finally, in order to investigate the effect of different variables on
the predicted spectropolarimetric signals we calculate a series of
ensembles holding all parameters constant except one. We refer to the
resulting set of ensembles as a ``run.'' To visualize the effects of
that parameter, in Figures \ref{f:pow-rcl-32} to \ref{f:pow-tcf-2.4} we plot the the Gaussian fit parameters and modes
determined for each metric in each ensemble versus the varyied
parameter.  The width of the Gaussian is shown by the gray area. We
plot the true mode for zero-centered metrics (\ie \SEW) and non-zero
mode for metrics that are not expected to be inherently symmetric.

\section{Results}
\label{s:results}
%

We have performed a preliminary survey of the effect of three
interrelated factors on the spectropolarimetric signatures of SNe:
clump size, number of clumps and TCF. We note that in addition to
likely parameter values, for some of these simulations we use numbers
or sizes of clumps that are at physical extremes in order to explore
the theoretical limits of the models. (For instance, an ensemble with
1024 clumps of radius \rvxl~would not be best characterized as
representing a ``clumpy ejecta,'' but can provide insight into trends
in the model. See Section \ref{ss:limits}.)

In this paper we present complete results for seven runs: two have
constant \rcl~and vary \ncl~(parameters in Table \ref{t:rcl}), two
have constant \ncl and vary \rcl (Table \ref{t:ncl}), and three vary
both \rcl and \ncl systematically to maintain an approximately
constant TCF (Table \ref{t:tcf}).  For each of these runs we show
trends for several metrics.

\subsection{Constant \rcl}
\label{ss:rcl}
%

We conducted two runs with constant clump radius, one with
\rcl=\rvxvi~and the other with \rcl=\rvxxxii~ (see Table \ref{t:rcl}
as well as Table \ref{t:model} for complete parameters).

\begin{figure}[htbp]
\begin{center}
\includegraphics[width=3in]{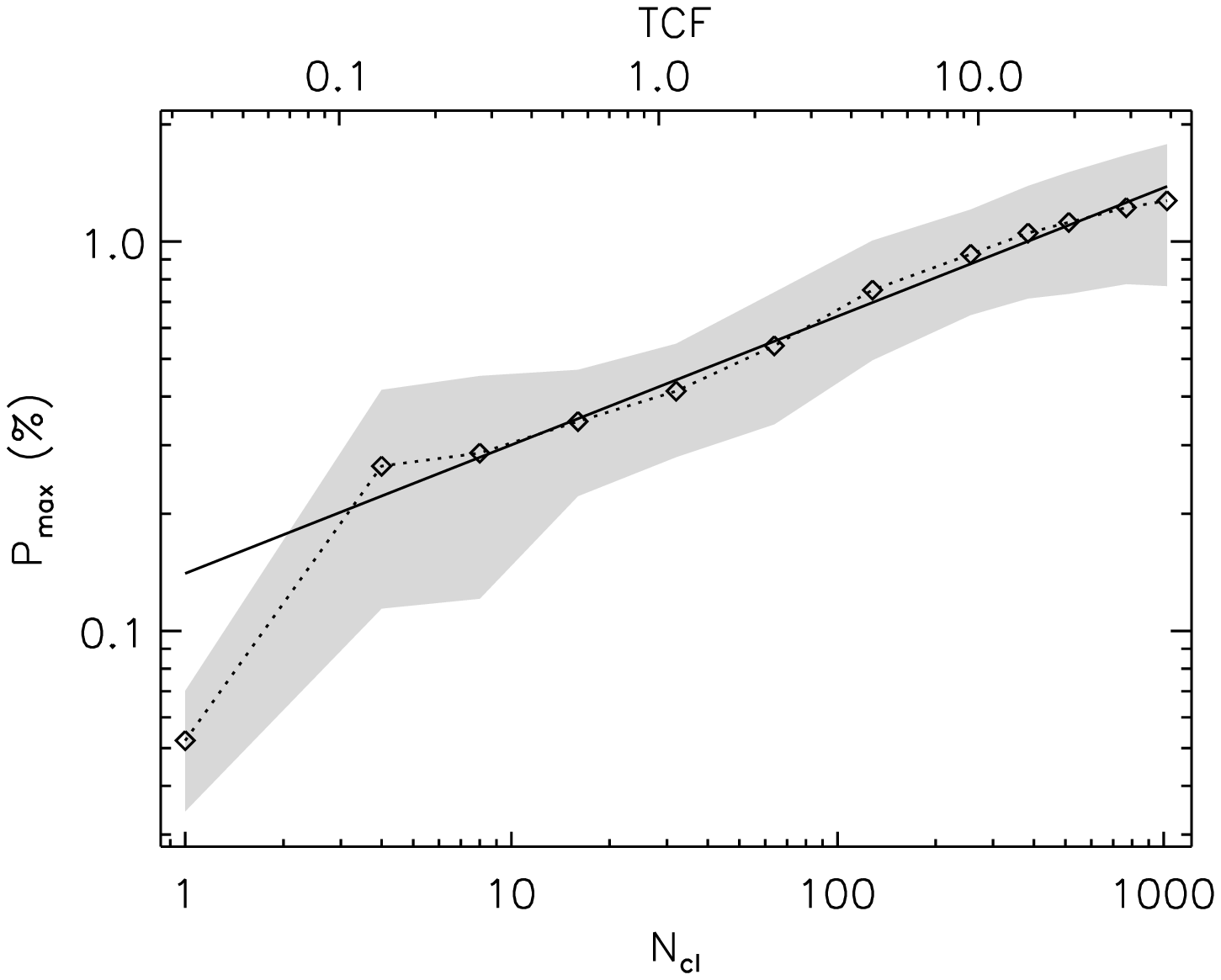}\\
\includegraphics[width=3in]{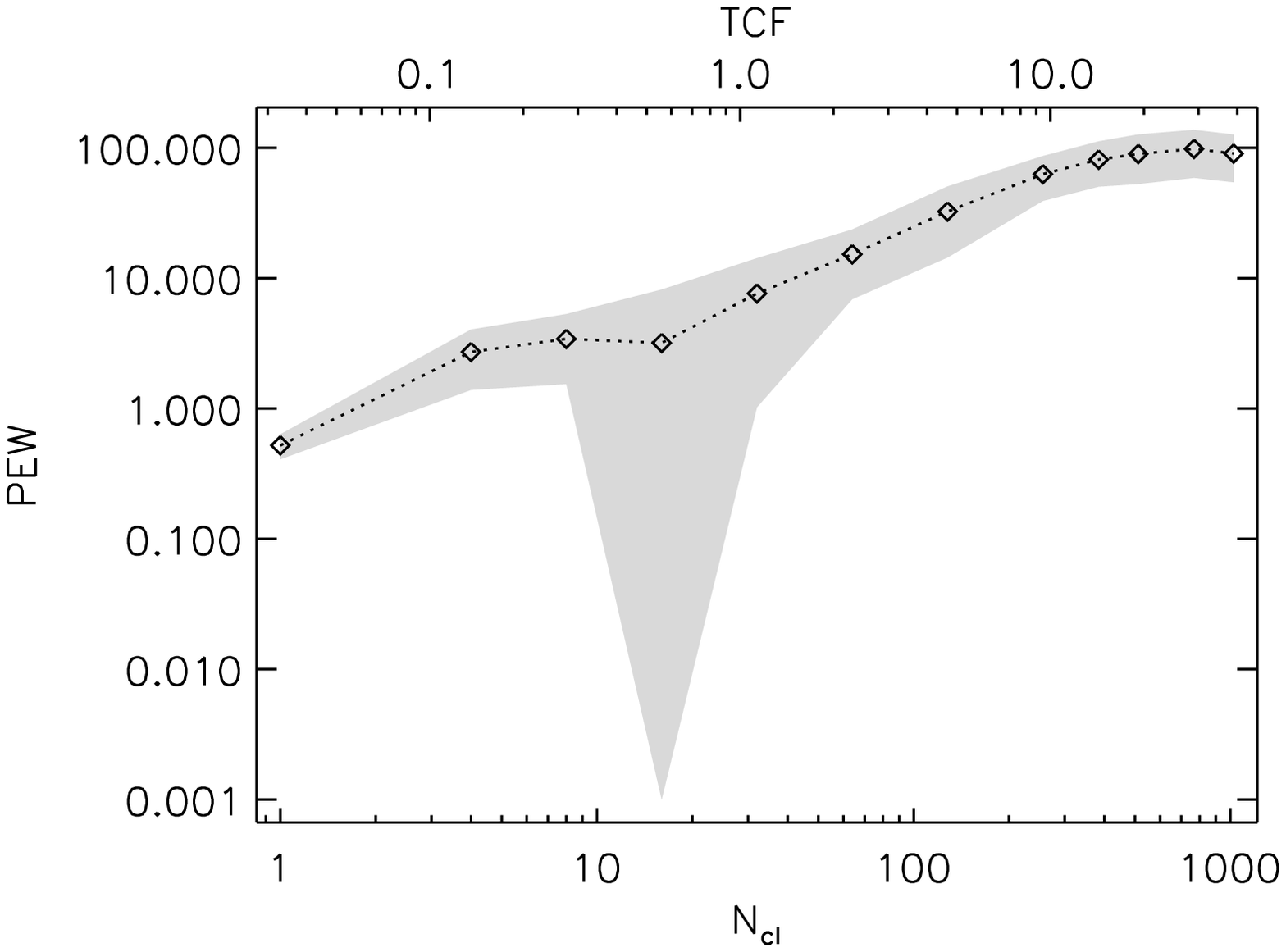}
\caption{The dependence of \MP~and PEW on \ncl~in the
\rcl=\rvxvi~run. A fit to the linear power-law portion (slope is 0.33) is
plotted in the \MP~figure (above).  TCF of the ensemble is shown, at the top of each figure.}
\label{f:pow-rcl-16}
\end{center}
\end{figure}

\begin{figure}[htbp]
\begin{center}
\includegraphics[width=3in]{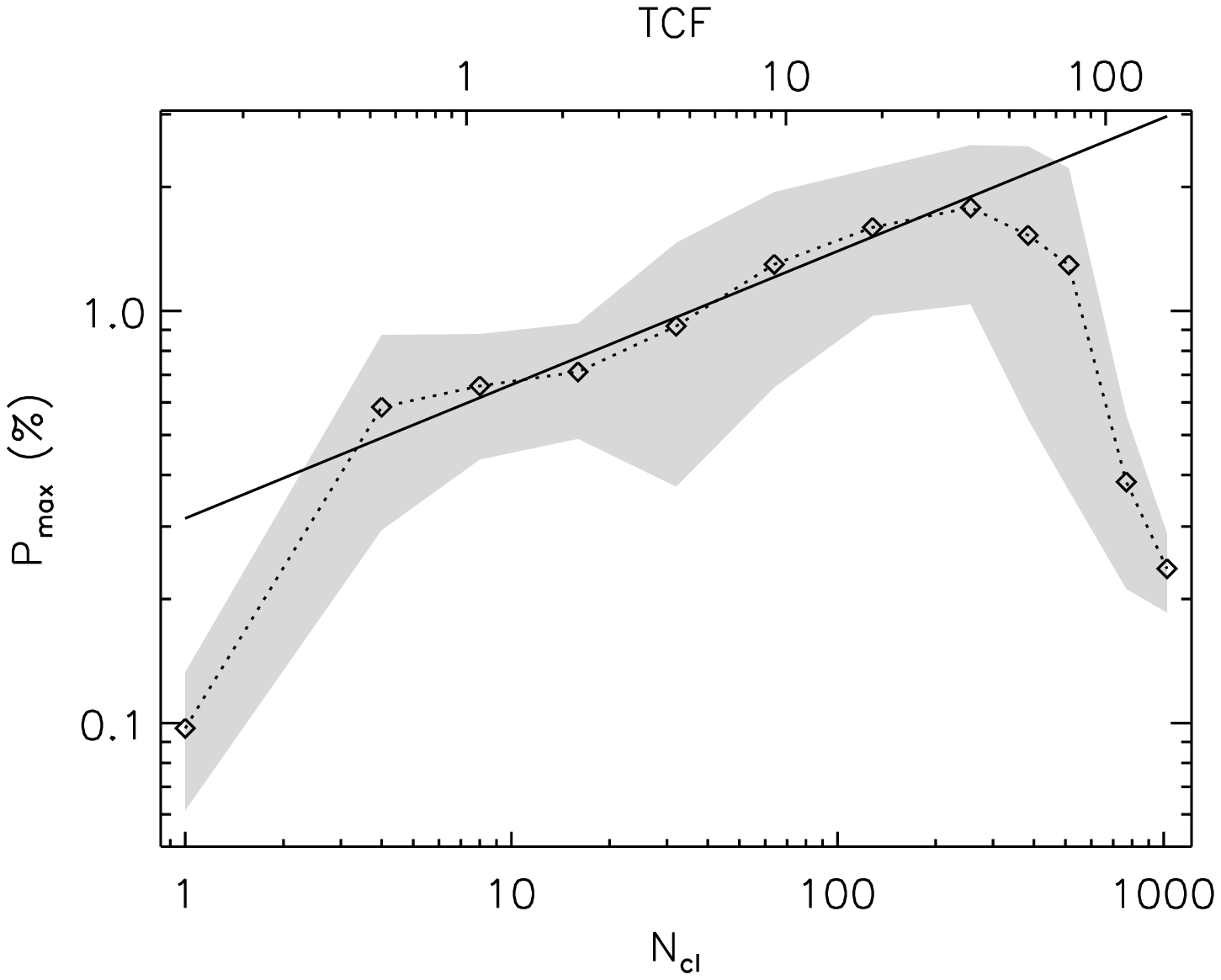}\\
\includegraphics[width=3in]{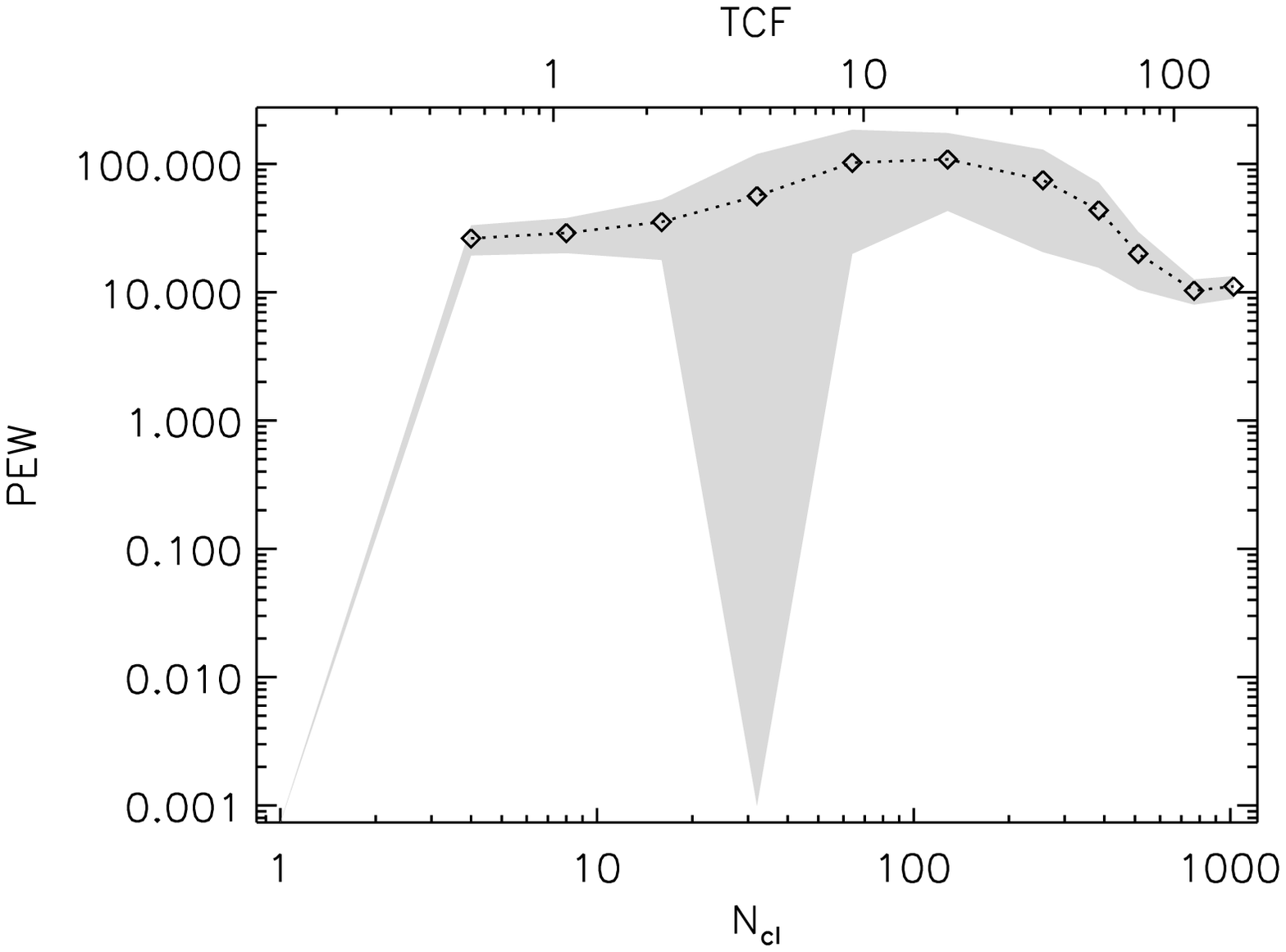}
\caption{The dependence of \MP~and PEW on \ncl~in the
\rcl=\rvxxxii~run. A fit to the linear power-law portion (slope is 0.32) is
plotted in the \MP~figure (above). Compared to the \rcl=\rvxvi~run
(Figure \ref{f:pow-rcl-16}) we see significant deviation from
power-law behavior at large numbers of clumps, indicating that
symmetric areas of the photosphere are being covered by the larger
clumps. TCF of the ensemble is shown, at the top of each figure.}
\label{f:pow-rcl-32}
\end{center}
\end{figure}

In our clump model, polarization of SNe is due to asymmetric obscuring
of the photosphere (see \S\ref{S:geometry}). We expect an increase in
polarization as the number of clumps increases. But beyond a certain
point, added clumps will create symmetries with other clumps, and the
polarization signal will decrease.  Therefore we begin by looking for
this pattern of increasing polarization at small \ncl~followed
decreasing polarization at large \ncl. While this behavior is not
readily apparent in the \rcl=\rvxvi~run (Figure \ref{f:pow-rcl-16}),
it is seen in the \rcl=\rvxxxii~series (Figure
\ref{f:pow-rcl-32}). This is likely due to the fact that larger clumps
will cover a broader range of Q and U vectors, meaning that it is
easier to regain symmetry with fewer clumps.

In the \MP~versus \ncl~plot of Figure \ref{f:pow-rcl-16} we have
overlaid a fit to the linear portion of the \MP~data, which has a
slope of 0.33. A similar set of plots for \rcl=\rvxxxii~are shown in
figure \ref{f:pow-rcl-32}; the slope of the central approximately
linear portion of the \MP~versus \ncl~plot is similar, 0.32.

\epsscale{1.}
\epsscale{1.}
\epsscale{1.}
\epsscale{1.}

\subsection{Constant \ncl}
\label{ss:ncl}

\epsscale{1.}
\begin{figure}[htbp]
\begin{center}
\includegraphics[width=3in]{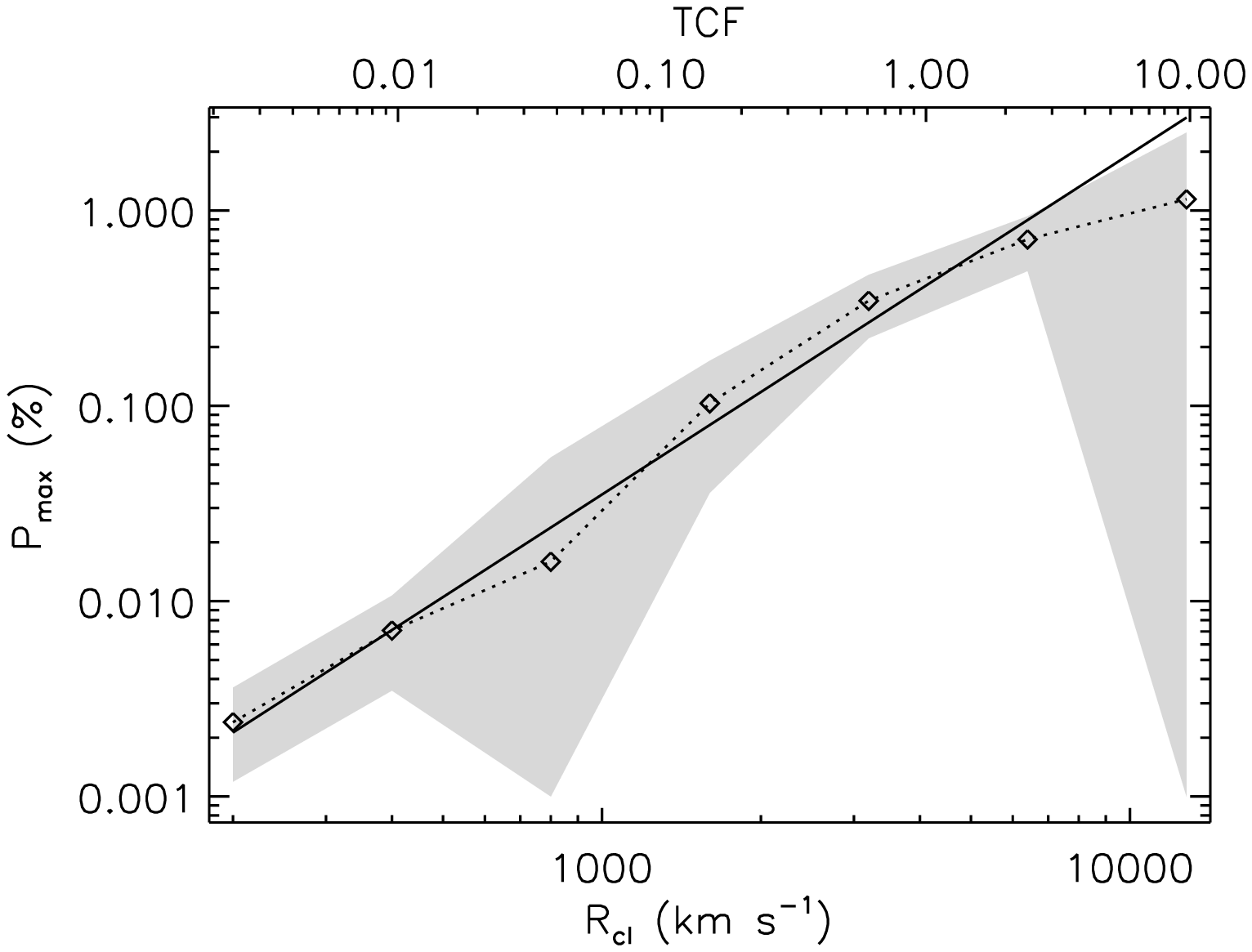}\\
\includegraphics[width=3in]{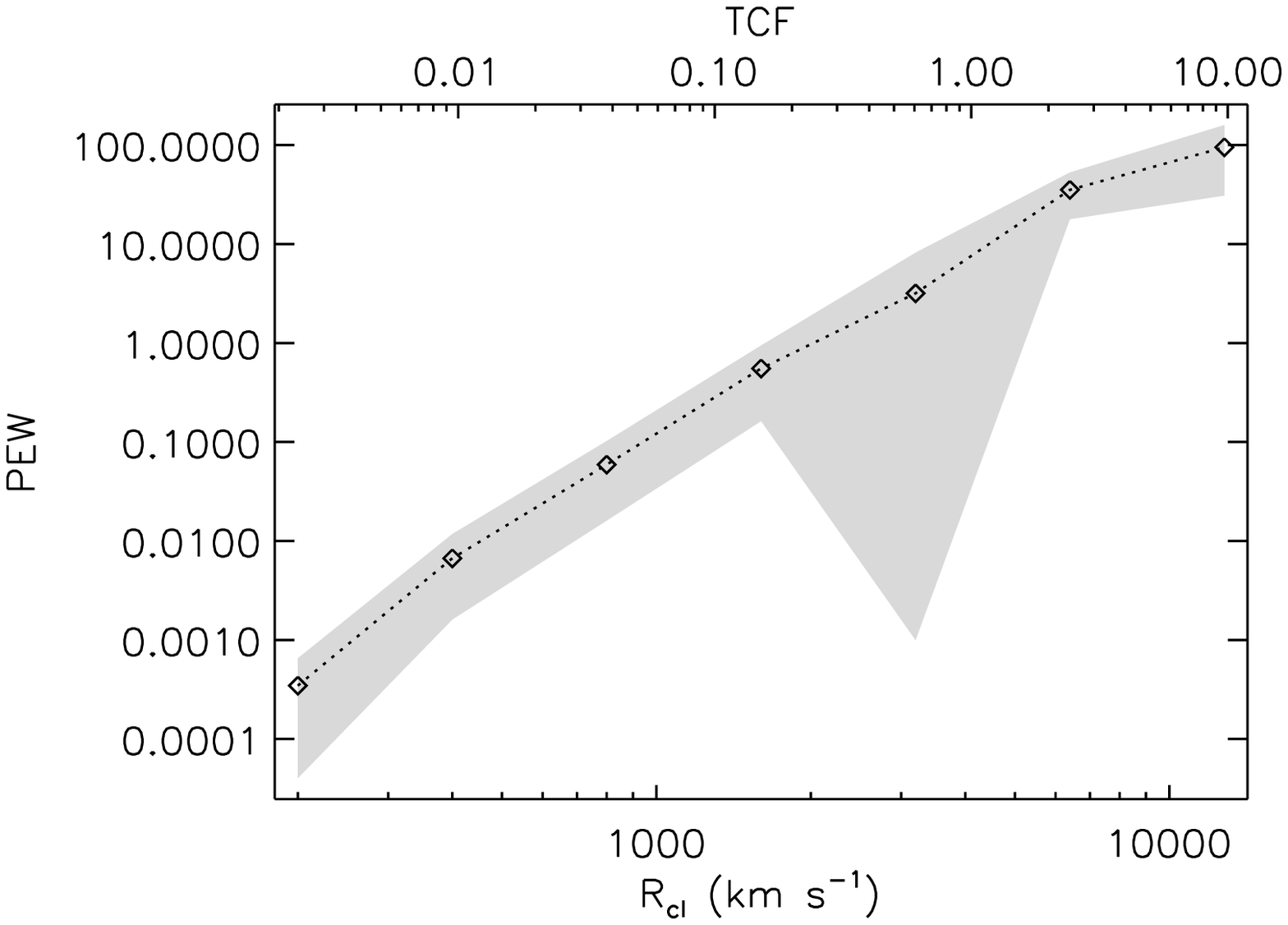}
\caption{The dependence of \MP~and PEW on \rcl~in the \ncl=16 run. A
fit to the linear power-law portion (slope is 1.74) is plotted in the
\MP~figure (above). TCF of the ensemble is shown, at the top of each figure.}
\label{f:pow-ncl-16}
\end{center}
\end{figure}
\epsscale{1.}

\epsscale{1.}
\begin{figure}[htbp]
\begin{center}
\includegraphics[width=3in]{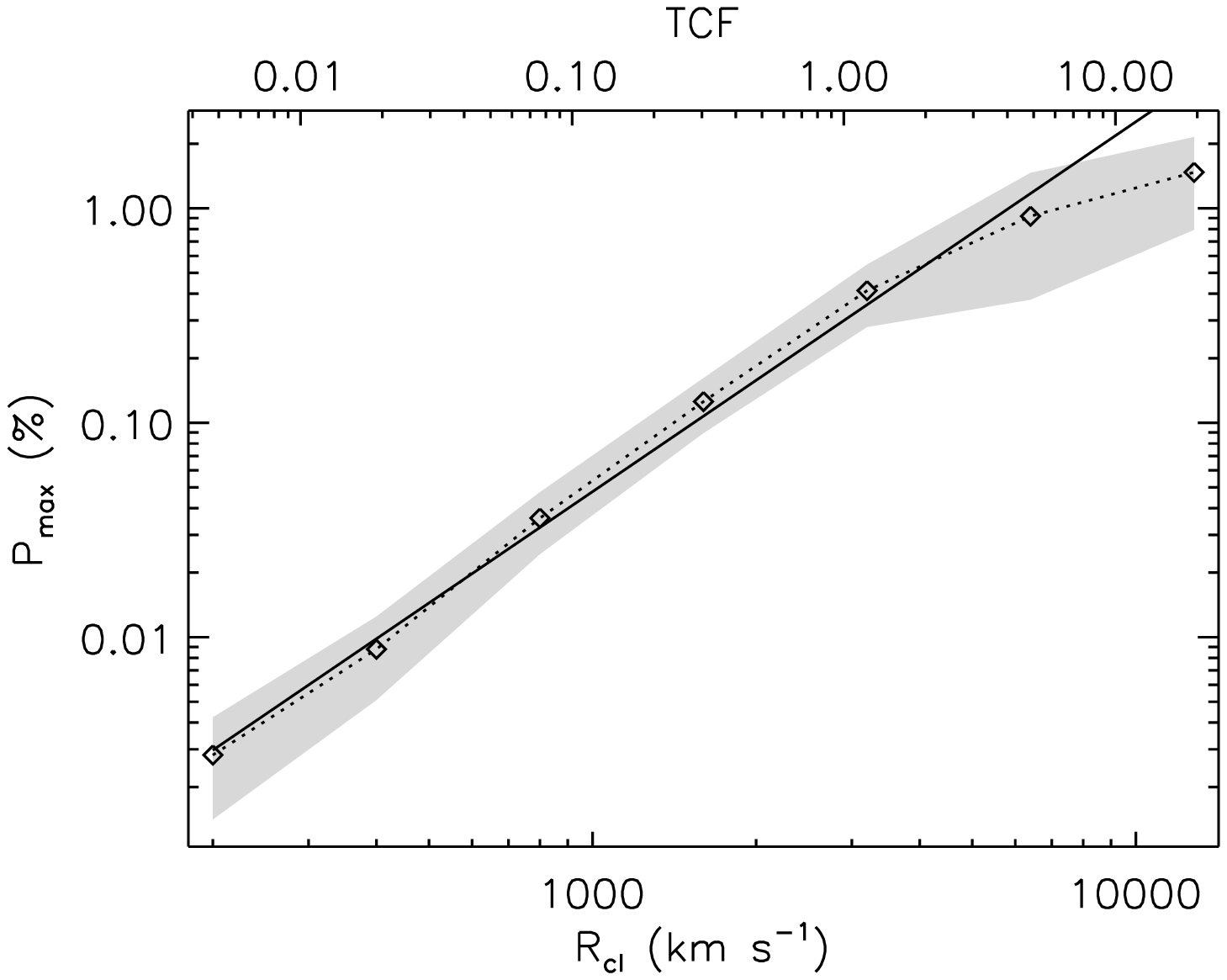}\\
\includegraphics[width=3in]{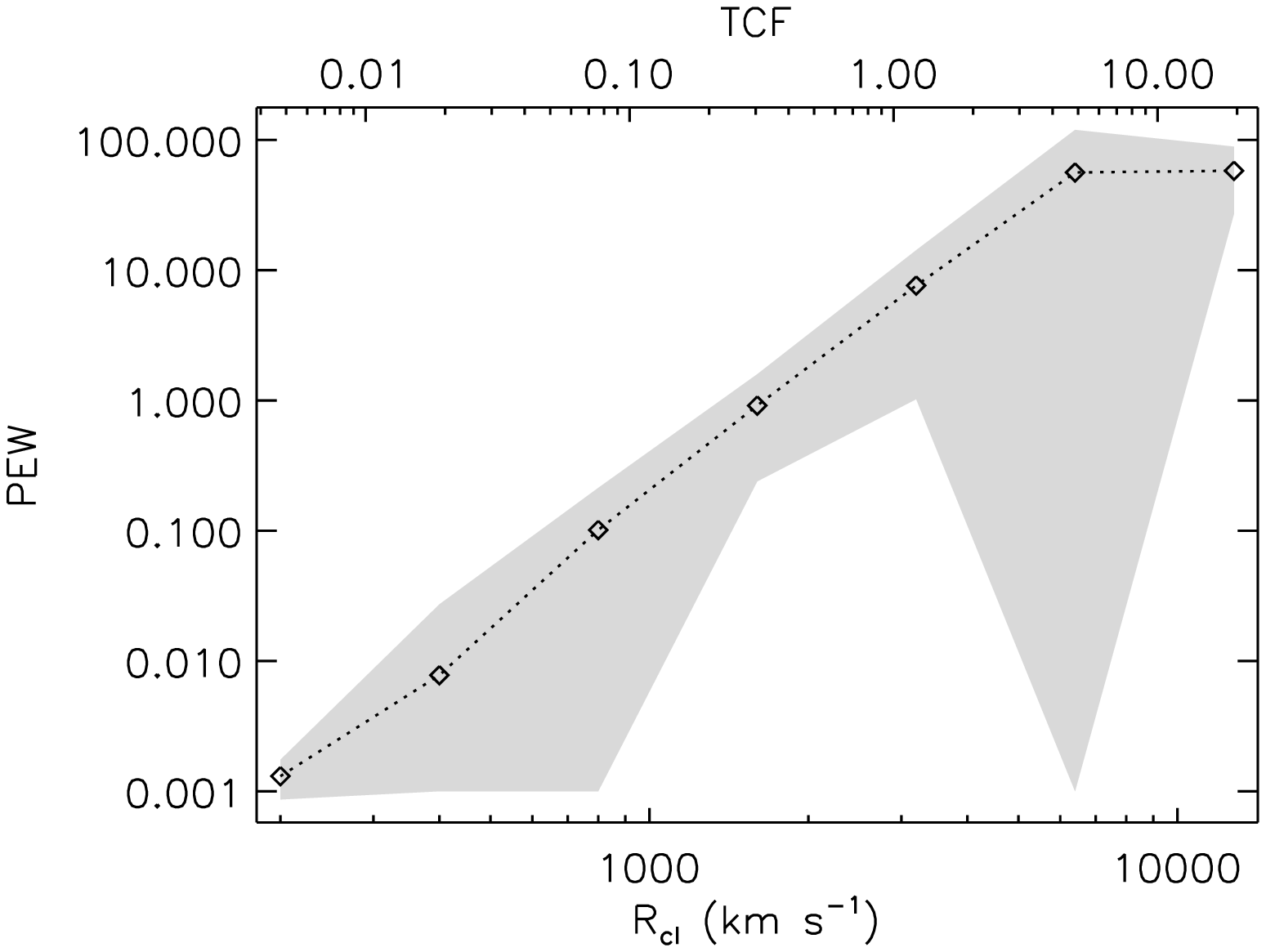}
\caption{The dependence of \MP~and PEW on \rcl~in the \ncl=32 run. A
fit to the linear power-law portion (slope is 1.72) is plotted in the
\MP~figure (above).  TCF of the ensemble is shown, at the top of each figure.}
\label{f:pow-ncl-32}
\end{center}
\end{figure}
\epsscale{1.}

We conducted two runs with a constant number of clumps, one with
$n_{\rm cl}=16$ and the other with \ncl=32 (see Table \ref{t:ncl} as
well as Table \ref{t:model} for complete parameters).

Figures \ref{f:pow-ncl-16} and \ref{f:pow-ncl-32} show that in the
constant \ncl~runs \MP$\propto$\rcl.  At larger \rcl~this relationship
breaks down due to increasing geometrical effects from clump overlap,
both spatially in the line-of-sight and spectrally in velocity space.
In the \MP~versus \rcl~plots of Figures \ref{f:pow-ncl-16} and
\ref{f:pow-ncl-32}, we have overlaid a fit to the linear portion of
the data. The slopes are 1.74 and 1.72, respectively.

\subsection{Constant TCF}
\label{ss:tcf}

\begin{figure}[htbp]
\begin{center}
\includegraphics[width=2.7in]{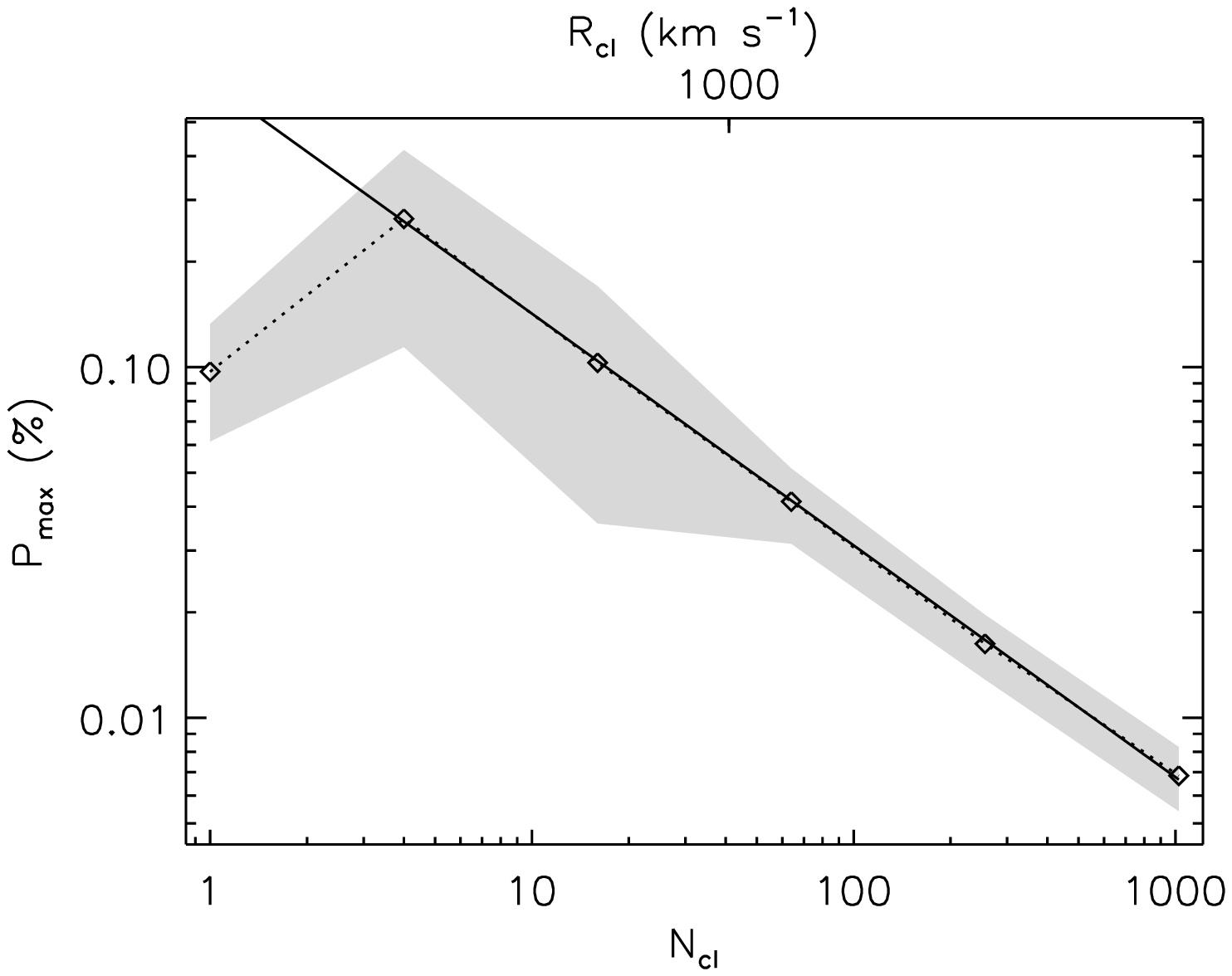}\\
\includegraphics[width=2.7in]{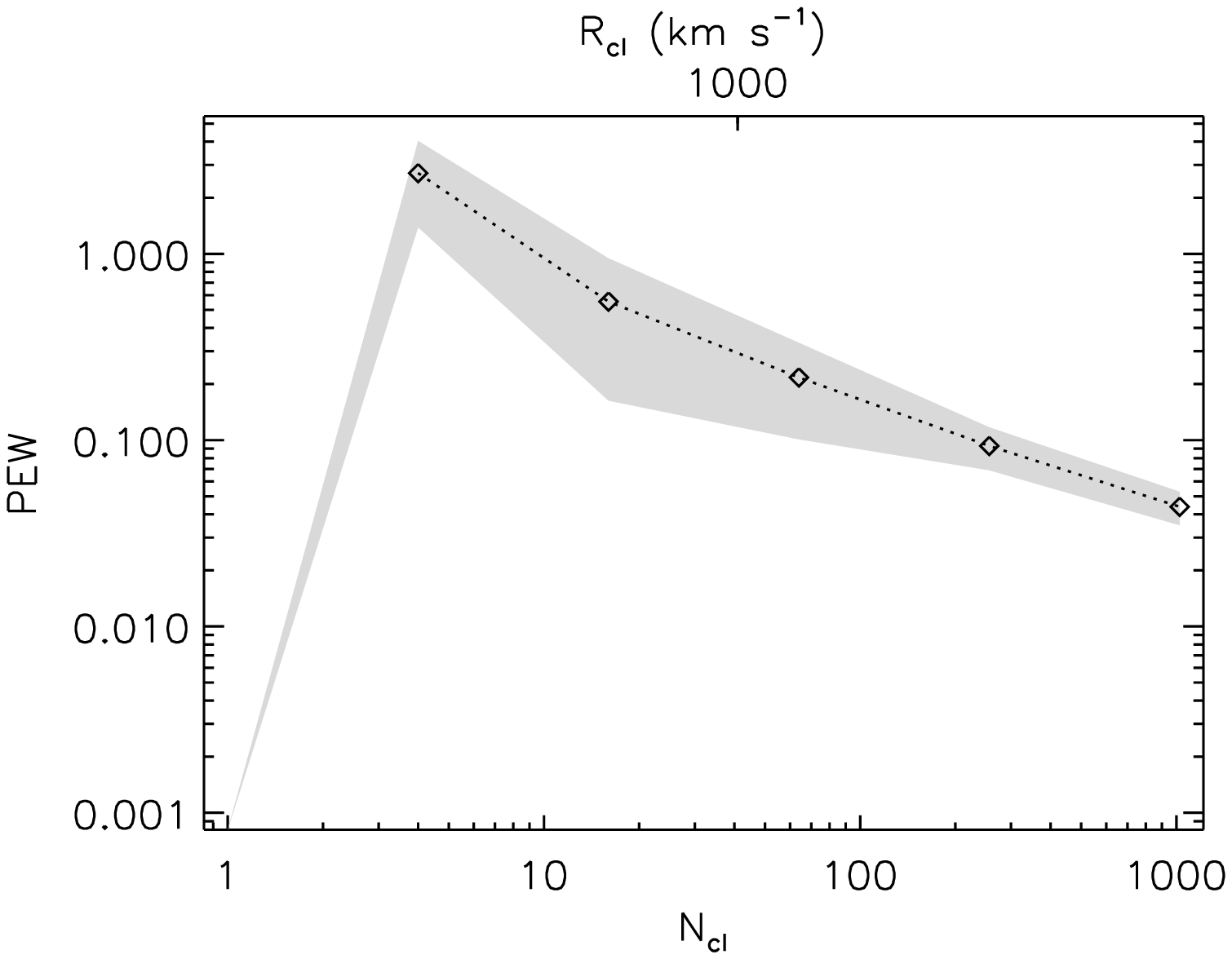}
\caption{The dependence of \MP~and PEW on \ncl~and \rcl~in the
TCF$\simeq0.15$ run. Linear fits are $\propto-0.66$\ncl~and
$\propto1.32$\rcl.}
\label{f:pow-tcf-15}
\end{center}
\end{figure}

\begin{figure}[htbp]
\begin{center}
\includegraphics[width=2.7in]{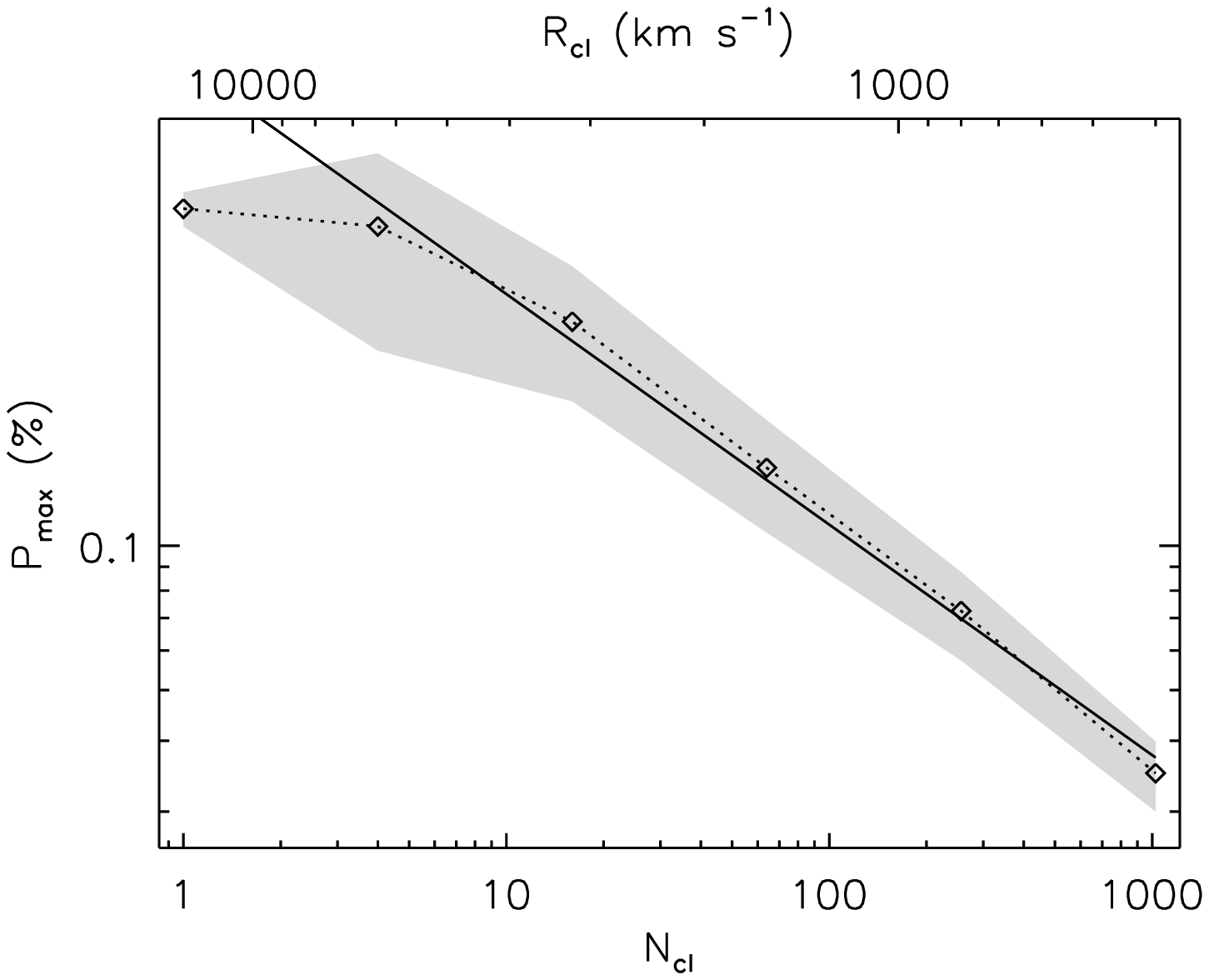}\\
\includegraphics[width=2.7in]{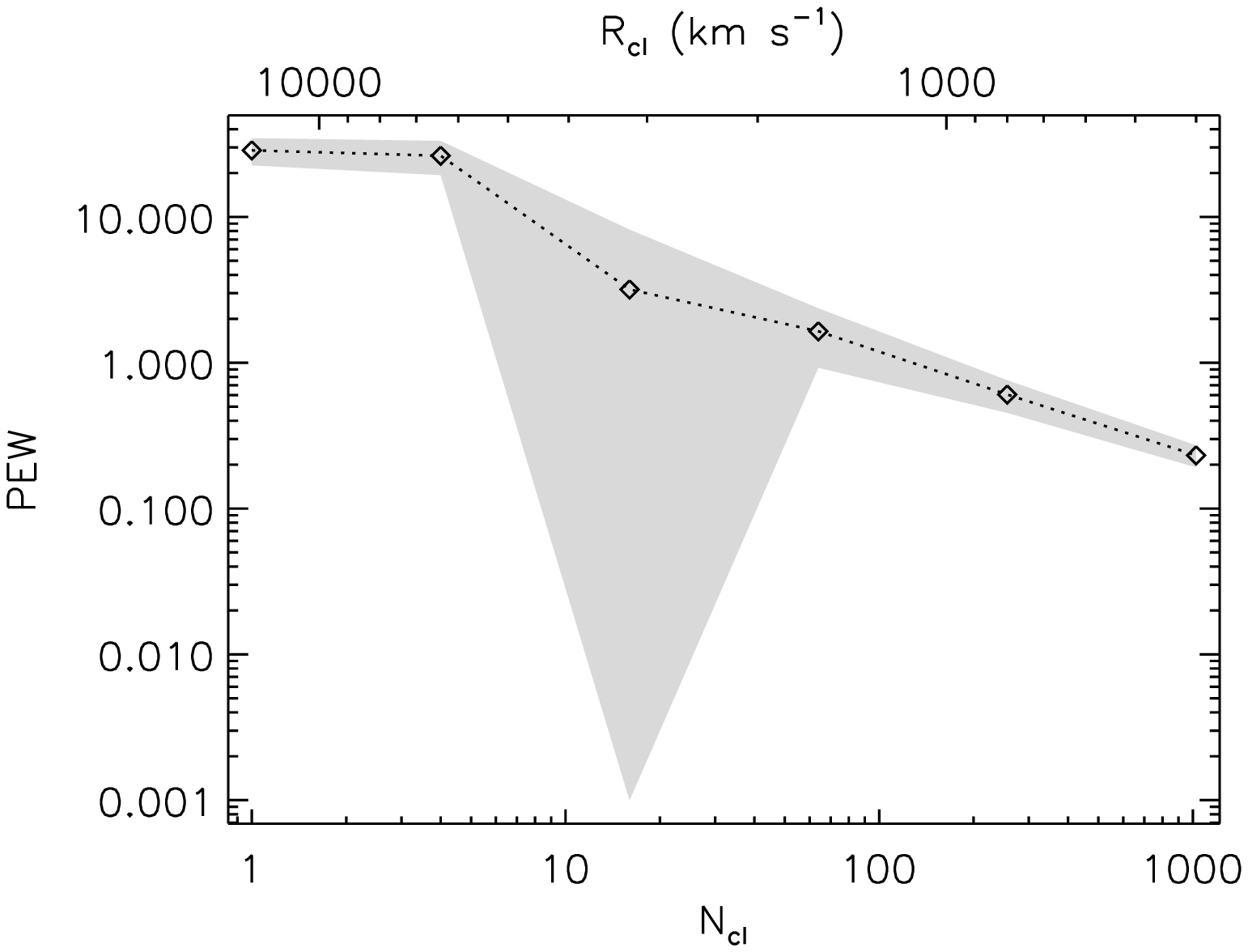}
\caption{The dependence of \MP~and PEW on \ncl~and \rcl~in the 
TCF$\simeq0.60$ run. Linear fits are $\propto-0.55$\ncl~and 
$\propto1.11$\rcl.} 
\label{f:pow-tcf-60}
\end{center}
\end{figure}

\begin{figure}[htbp]
\begin{center}
\includegraphics[width=2.7in]{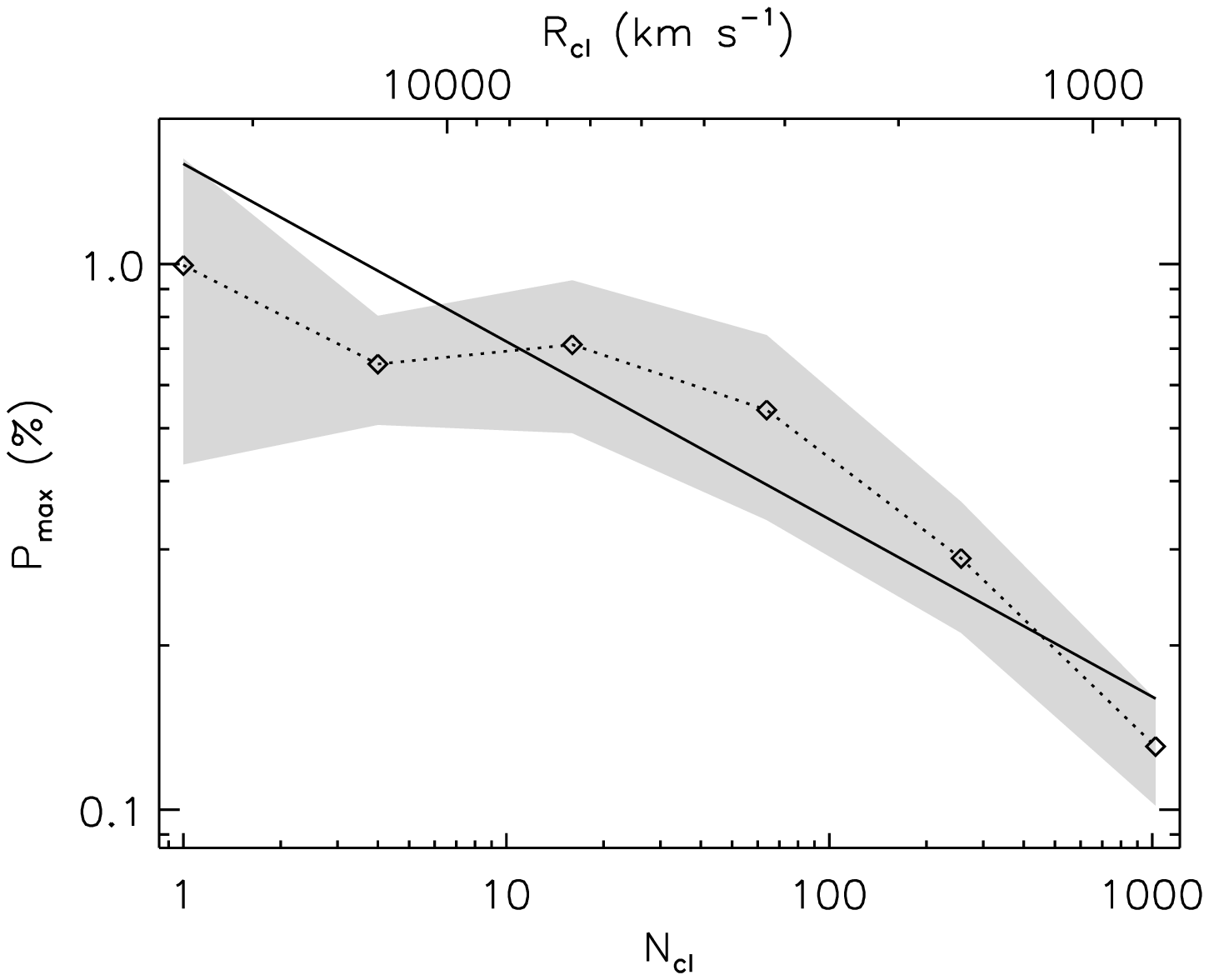}\\
\includegraphics[width=2.7in]{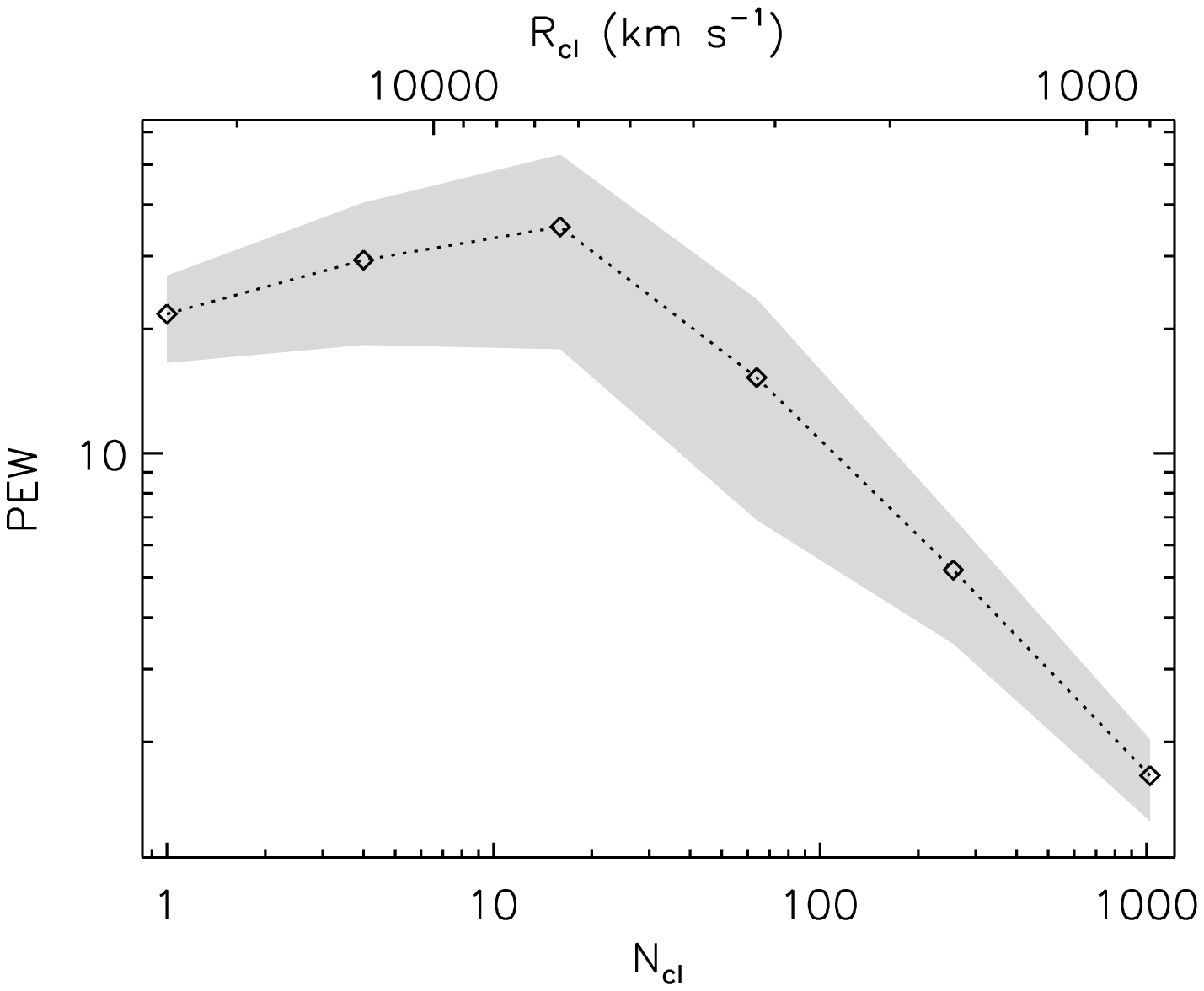}
\caption{The dependence of \MP~and PEW on \ncl~and \rcl~in the 
TCF$\simeq2.4$ run. Linear fits are $\propto-0.33$\ncl~and 
$\propto0.65$\rcl.} 
\label{f:pow-tcf-2.4}
\end{center}
\end{figure}

For this series of runs, we held the theoretical covering fraction
(TCF, see \S \ref{s:anal_ensem} for definition) constant, varying \ncl
from 1 to 1024 and \rcl~proportionately to maintain a TCF of
$\approx0.15$, $\approx0.60$ and $\approx2.4$. The actual TCF obtained
for each ensemble is closest to the nominal value at higher \ncl~where
the value is less likely to be skewed by random placements of
clumps. Figures \ref{f:pow-tcf-15}, \ref{f:pow-tcf-60} and
\ref{f:pow-tcf-2.4} all show that, as anticipated, large numbers of
small clumps leads to recapturing symmetry and a decrease in
polarization.

In Figures \ref{f:pow-tcf-15}, \ref{f:pow-tcf-60} and
\ref{f:pow-tcf-2.4} we show the log-log plots for the
TCF~$\simeq~0.15$, TCF~$\simeq~0.60$ and TCF~$\simeq~2.4$,
respectively. In these plots, we have overlaid fits to the linear
portions of both \MP~plots. In the TCF $\simeq0.15$ case, we find that
this slope is 1.32 in the \rcl~plot and -0.66 in the \ncl~version; for
TCF $\simeq0.60$ they are 1.11 and -0.55 respectively. In the
$\simeq2.4$ run, a fit to the data gives slopes of 0.65 for \rcl~and -0.33 for \ncl, though the behavior no longer appear to be truly linear, due to complicated effects of overlapping of clumps.

In the constant \rcl~and \ncl~runs (\S \ref{ss:rcl} and \S
\ref{ss:ncl}), the slope of the fit to the \MP\ remained the same
between runs, at least in the portions that are roughly linear. The
TCF runs show less uniformity (even excluding the highest TCF run
where behavior does not seem linear) than in the other cases. This may indicate a trend
where increased covering fraction leads to a flattening of
polarization dependence in both \rcl~and \ncl.

The difference between the TCF $\simeq0.15$ and TCF $\simeq0.60$ is
likely due to small statistical variations at very small covering
fraction. We therefore take the results from the TCF $\simeq0.60$ run
as most representative of the regime of power-law behavior for the TCF
runs, where there are enough clumps for reasonable statistics, but not
so many that complex overlapping effects make the polarization
dependence non-linear.

\subsection{Polarization dependence on \ncl~and \rcl}

In order to understand these results, we seek a simple model for the
relationship between polarization and \ncl, \rcl~and TCF.  We
therefore propose the following simple analytical model and compare it
to our model's numerical behavior in order to test consistency.
and to provide insight into both simple and complex regimes.

We begin our proposed model with the polarization due to one clump,
$P_{\rm cl}$. The polarization from that clump will depend on the
photospheric area covered by the clump, $\propto r_{\rm cl}^{2}$. But
the polarization will also depend on what portion of the photosphere
is covered -- if the clump is very large, it will eventually cover
orthogonal vectors in Q-U space, and therefore reduce the polarization
due to that clump. Thus we make the assumption that the polarization
due to the clump will be proportional to the area of the clump times
an unknown function of \rcl, which we in turn assume to be a power of
\rcl:
\begin{equation} \label{eq:pcl}
P_{\rm cl} \propto  r_{\rm cl}^{2} f(r_{\rm cl})
 \propto  r_{\rm cl}^{2} r_{\rm cl}^m
\end{equation}

The polarization should also be a function of \ncl, which should be
roughly linear for small covering fractions.  
As we increase the number of clumps, we will begin to have more than one clump at the same velocity, each covering different Q and U vectors. 
While $\langle Q \rangle = 0$,  $\langle Q^2 \rangle$ is not. Like a random walk, $Q^2 \propto n_{\rm cl}$. The same is true for U. Thus, since $P = \sqrt{Q^2+U^2}$ (equation \ref{eq:p}), the total polarization will also increase as $\sqrt{n_{\rm cl}}$. (A similar derivation of the $P \propto \sqrt{n_{\rm cl}}$ behavior is shown in \citet{richardson96}.)
Thus in our model, the total polarization will depend on the number of
clumps and the polarization per clump:
\begin{equation} \label{eq:ptot1}
P_{tot} \propto g(n_{\rm cl}) P_{\rm cl} \sqrt{n_{\rm cl}}
\end{equation}

Here $g(n_{\rm cl})$ is an as-yet unspecified function of
\ncl~representing the non-linear effects on the total polarization due
to multiple clumps.  Assuming that $g(n_{\rm cl}) \propto n_{\rm
cl}^l$ and combining equation \ref{eq:ptot1} with equation
\ref{eq:pcl} this becomes 
\begin{equation} \label{eq:ptot}
P_{tot} \propto n_{\rm cl}^l r_{\rm cl}^m n_{\rm cl}^{1/2} r_{\rm cl}^{2} =
n_{\rm cl}^{l+1/2} r_{\rm cl}^{m+2} 
\end{equation}

Equation \ref{eq:ptot} predicts that within a run, the total
polarization should have a power-law dependence on \rcl~and \ncl. This
prediction agrees with the behavior of our models shown in
\S\ref{ss:rcl} and \S\ref{ss:ncl}. The actual dependence can be
disentangled by holding either \rcl~or \ncl~constant:
\begin{equation} \label{eq:constpow}
P_{tot} \propto \left\{ \begin{array}{rcl}

r_{\rm cl}^{m+2} & ~ & \mbox{ for const. \ncl~} \\
n_{\rm cl}^{l+1/2} & ~ & \mbox{ for const. \rcl~}
\end{array} \right.
\end{equation}

We can now use our simulations to determine the values of $l$ and $m$.
The metric that will most directly correspond to the $P_{tot}$ derived
above will be \MP~ because it is less influenced by complicated
geometry effects in the line. (PEW~should also follow this
relationship in cases where polarization is due to a few discrete
clumps.)

From our simulations, the slope of \MP~as a function of \ncl~for
constant \rcl~is 0.33 (Figure \ref{f:pow-rcl-16}). Thus $n_{\rm
cl}^{l+1/2}= n_{\rm cl}^{0.33}$, and $l=-0.17$. Our results for
constant \ncl~(Figure \ref{f:pow-ncl-16}) show \MP~as a function of
\rcl~is $\propto r_{\rm cl}^{1.74}=r_{\rm cl}^{m+2}$, and so
$m=-0.26$.

To check for consistency, we can use our model and these results to
predict the behavior of \MP~for runs with constant TCF. For constant TCF,
$n_{\rm cl} r_{\rm cl}^2$ is constant, and $n_{\rm cl} \propto r_{\rm
cl}^{-2}$ and $r_{\rm cl} \propto n_{\rm cl}^{-1/2}$. Substituting
these relations into eqn. \ref{eq:ptot} we find
\begin{equation} \label{eq:tcfpow}
P_{tot} \propto  \left\{ \begin{array}{l}
 r_{\rm cl}^{1.08} \\ 
 n_{\rm cl}^{-0.54}
\end{array} 
\right\} 
\mbox{const. TCF }
\end{equation}

These predicted slopes agree remarkably well with the fit to our
numerical results: $ r_{\rm cl}^{1.11} $ and $ n_{\rm cl}^{-0.55}$
(Figure \ref{f:pow-tcf-60}).

Note that this implies that the total polarization is decreased by
\ncl$^{-0.17}$ if there is substantial clumping. Similarly, total polarization is decreased by \rcl$^{-0.26}$. This 
implies that the larger the radius of the clump, the more likely it is
to extend into less polarized parts of the photosphere or cover
complementary Q and U vectors, and thereby reduce the resulting
polarization per clump.

Similar behavior for the polarization can be derived by introducing clumps to the integrals in equations \ref{eq:dk1} and \ref{eq:dk2}. For non-overlapping, optically thick clumps, the average polarization would be propto $n_{\rm cl}^{-1/2}$ and propto $r_{\rm cl}^{1}$. The difference from these values in our results is due to the more complicated effects, such as overlap and velocity-width of the clumps, that can be handled more effectively in the numerical model.

\subsection{Connection to Observations}

One of the greatest obstacles to the effective use of
spectropolarimetry in astrophysics is the degeneracy between source
and signal -- because polarization is a second order effect, there
will likely be more than one host configuration that will produce a
given observation. One of the most important goals of this paper is to
find ways to break polarimetric degeneracy where possible, and
otherwise to estimate the range of possible configurations leading to
an observation, and their relative likelihoods.

Two factors that can break degeneracy in polarimetry are wavelength
and time dependence. For instance, while continuum polarization for an
ellipsoidal photosphere may be the same as that for a spheroidal
photosphere with chemically inhomogeneous ejecta, the wavelength
dependence of the polarization will be different.  The former case
will also likely maintain the same degree of polarization and position
angle over time, while in the latter, individual clumps will
continually become optically thin over time as the ejecta expands,
likely resulting in variation in both degree and orientation of
polarization.

Our code incorporates the first of these factors intrinsically, namely
wavelength dependence in the line profile. Degeneracies for a single
SN can be further constrained using observations at different
epochs. Take, for instance, the case where one epoch of observations
indicate that the asymmetry may be due to a single large clump
covering most of the photosphere, or to several overlapping
medium-sized clumps at about the same distance from the
photosphere. As the ejecta expands, the outer portions of the clumps
will become optically thin while the portion of each clump closest to
the photosphere may remain optically thick. In later observations of
the single-clump case, we would observe the signature of one clump
reducing in size. With several overlapping clumps, as we see further
towards the center of the SN the single clump would appear to break up
and the spectropolarimetric signature would become that of several
small clumps. Our simulations predict that this difference should be
detected in the relative change in \MP~and PEW~at different epochs.

\begin{figure}[htbp]
\begin{center} \plotone{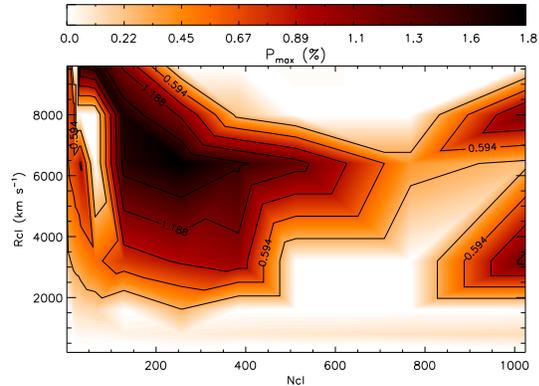}
\caption{The expectation value of \MP~as a function of \ncl~and
\rcl~predicted by our simulations. Here, a measurement of \MP~in a SN
line would be a plane, constraining the host system's parameters to
those described by an intersecting contour in the above plot.  }
\label{f:mp-cont}
\end{center}
\end{figure} 

\begin{figure}[htbp]
\begin{center}
\plotone{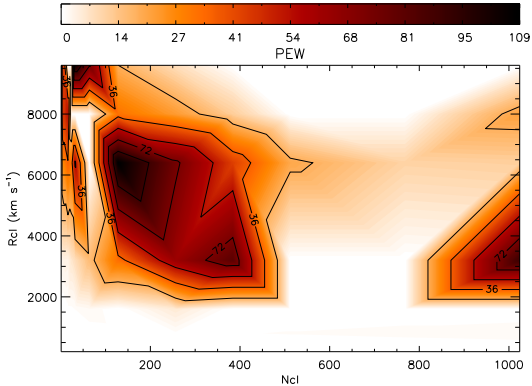}
\caption{The expectation value of PEW in the line as a function of
\ncl~and \rcl~predicted by our simulations. Here, a measurement of PEW
in a SN line would be a plane, constraining the host system's
parameters to those described by an intersecting contour in the above
plot. See the caption of Figure \ref{f:mp-cont} for a description of
the dashed and cross-hatched regions.}
\label{f:fd-cont}
\end{center}
\end{figure}

More directly, our simulations also give us a powerful method of
constraining clump parameters by using observational metrics from the
same epoch. Using our results from across all of our runs, we can
create a map of a metric, such as \MP, as a function of \ncl~and
\rcl~(Figure \ref{f:mp-cont}). This map will have the form of a
topographical map, though rather than having a single value at each
point, the surface would have a thickness representing the likely
range of \MP~from our Gaussian fit for that (\ncl,\rcl). An
observational measurement of \MP~in the line would be a plane, also
with a thickness representing the uncertainty in the measurement. That
plane would intersect our map, constraining the source \ncl~and
\rcl~to the contour thus created. The intersection of uncertainties
from our model and observations would give an uncertainty on the
prediction.

This method becomes even more powerful when we use more than one
metric: the intersection of observed \PEW~with our model (Figure
\ref{f:fd-cont}) would create another contour in \rcl~and \ncl, and
the two contours could be combined to further constrain source
geometry. Each metric would provide another constraint and narrow the
possible source parameters.  Consistency between these predictions for
a given observation also provides a test of how well our models
represent the physical system of the SN.

\section{Summary}
\label{s:summary}
%

In this paper, we have presented a new spectropolarimetric radiative
transfer code, designed to model the effects of chemical
inhomogeneities (``clumps'') in the ejecta of SNe on a pure-scattering
line profile. While most approaches to the modeling of
spectropolarimetry have focused on detailed modeling of a small number
of host geometries \citep[\eg][]{hoeflich09,kasen06} we have instead chosen to use simplified radiative
transfer to model large numbers of host configurations (realizations)
with a broad range of parameters. This approach has two goals: first,
to create a catalog of spectropolarimetric line profiles and host
geometries that can be compared to observations; and second to give an
estimate of how unique (or degenerate) is the connection between model
and observation.

This paper presents the results of our first survey of the likely
polarization signals from clumpy SN ejecta, specifically the effects
of the number and size of clumps (and the related effect of covering
fraction) on the emergent spectrum and Stokes vectors. We have paid
special attention to the connection between underlying clump geometry
and predictions for robust observational parameters, such as \MP~and
\PEW~of the line. By themselves, these results allow observers to
understand the range of possible configurations that may produce a
given measurement, including the relative likelihood of each.

One important result of our simulations is that random clumpiness can
regularly produce peak line polarization of 1\% with small to medium clumps, and
up to 1.6\% in simulations where we allow clumps to largely obscure
the photosphere at some wavelengths (see \ref{ss:limits}); polarization produced 
by individual realizations can produce even higher values.  These average values are in
the range observed for SNe Type Ia (1-2\%), and indicate that
line polarization is a plausible explanation for existing data. 

We have also shown that loops in the Q-U plane, which are often
observed in SNe, can be produced by clumpiness alone.  \citet{kasen03}
showed that such loops could be produced by a clump above an
aspherical photosphere.  Here we have shown that these loops can be
produced with a spherical photosphere but multiple clumps, if clumps with different position angles overlap in velocity (see figures \ref{f:spec_lots} and \ref{f:spec_loop}).

In order to connect our simulations to observations more concretely we
present the results of our models for observable metrics \MP~in the
line and PEW~as a function of \ncl~and \rcl. A measurement of \MP~is a
flat plane with finite thickness given by observational uncertainty,
and the intersection of plane and model creates a locus of possible
\rcl~ and \ncl~values in the host that could have led to the
observations. Combining results for multiple metrics further restricts
the possible ranges of \rcl~and \ncl. This method provides good
constraints on host geometry from even a single epoch of data.  Using
multiple epochs of data for the same supernova, the possible host
configurations can be constrained even further.  In the future we will
investigate other observables to find other independent constraints
that may produce a unique \rcl-\ncl~estimate. The intersection of
multiple metrics should also provide an indication of the precision
and accuracy of our maps.
 
The results presented in this paper show the potential for our
approach in giving insight into the complicated dependence of
polarization on SN geometry. It also provides a connection between the
theoretical emergent Stokes spectra and robust observables, such as
\MP~and \PEW.

In the future, we will use this code to explore a larger range of
parameter space to predict the observational signature of more host
configurations. Once a library of configurations and
spectropolarimetric signatures has been developed, we will be able to
understand degeneracies in the effects of a variety of host
parameters, as well as develop methods to break these degeneracies
with observations. Our simulations will also provide new observational
tests for SN explosion codes by predicting spectropolarimetric
signatures using characteristic clump sizes and distributions that
emerge from these models.

\acknowledgements KTH is grateful to Jay Gallagher, Jennifer Hoffman
and Rico Ignace for helpful discussion on this work. This research was funded in part by NSF grants 
AST-0922981, PHY 05-51164 and AST 07-07633, and by the DOE SciDAC Program (DE-FC02-06ER41438).


\bibliographystyle{apj}     
\bibliography{apj-jour,sne}



\begin{table}
\begin{center}
\caption{Model parameters for investigation the effect of varying the
number of clumps while keeping the clump size constant. \label{t:rcl}
\vspace{.1cm}}
\begin{tabular}{ccc}
\multicolumn{3}{c}{Run: $r_{\rm cl}=0.16 v_{max}$}\\
\tableline\tableline
$n_{\rm cl}$ & $r_{\rm cl}$ & TCF\\
 & \kms &\\
\tableline
1 & 3,200 & 0.02\\
4 & 3,200 & 0.14\\
8 & 3,200 & 0.28\\
16 & 3,200 & 0.61\\
32 & 3,200 & 1.23\\
64 & 3,200 & 2.24\\
126 & 3,200 & 4.87\\
256 & 3,200 & 9.76\\
384 & 3,200 & 14.61\\
512 & 3,200 & 19.42\\
768 & 3,200 & 29.32\\
1024 & 3,200 & 38.85\\
\tableline
\end{tabular}
\hspace{1cm}
\begin{tabular}{ccc}
\multicolumn{3}{c}{Run: $r_{\rm cl}=0.32  v_{max}$}\\
\tableline\tableline
$n_{\rm cl}$ & $r_{\rm cl}$ & TCF\\
 & \kms &\\
\tableline
1 & 6,400 & 0.08\\
4 & 6,400 & 0.54\\
8 & 6,400 & 1.11\\
16 & 6,400 & 2.42\\
32 & 6,400 & 4.90\\
64 & 6,400 & 9.74\\
126 & 6,400 & 19.46\\
256 & 6,400 & 39.05\\
384 & 6,400 & 58.45\\
512 & 6,400 & 77.70\\
768 & 6,400 & 117.30\\
1024 & 6,400 & 155.42\\
\tableline
\end{tabular}

\end{center}
\end{table}


\begin{table}
\begin{center}
\caption{Model parameters for investigation the effect of varying the
clump radius while keeping the number of clumps constant. \label{t:ncl}
\vspace{.1cm}}
\begin{tabular}{ccc}
\multicolumn{3}{c}{Run: $n_{\rm cl}=16$}\\
\tableline\tableline
$n_{\rm cl}$ & $r_{\rm cl}$ & TCF\\
 & \kms &\\
\tableline
16 & 200 & 0.002\\
16 & 400 & 0.009\\
16 & 800 & 0.038\\
16 & 1,600 & 0.151\\
16 & 3,200 & 0.607\\
16 & 6,400 & 2.429\\
16 & 12,800 & 9.716\\
\tableline
\end{tabular}
\hspace{1cm}
\begin{tabular}{ccc}
\multicolumn{3}{c}{Run: $n_{\rm cl}=32$}\\
\tableline\tableline
$n_{\rm cl}$ & $r_{\rm cl}$ & TCF\\
 & \kms &\\
\tableline
32 & 200 & 0.005\\
32 & 400 & 0.019\\
32 & 800 & 0.077\\
32 & 1,600 & 0.306\\
32 & 3,200 & 1.225\\
32 & 6,400 & 4.900\\
32 & 12,800 & 19.602\\
\tableline
\end{tabular}

\end{center}
\end{table}


\begin{table}
\begin{center}
\caption{Model parameters for investigation the effect of varying the
number and radius of clumps to keep TCF constant. \label{t:tcf}
\vspace{.1cm}}
\begin{tabular}{ccc}
\multicolumn{3}{c}{Run: TCF $\simeq0.15$}\\
\tableline\tableline
$n_{\rm cl}$ & $r_{\rm cl}$ & TCF\\
 & \kms &\\
\tableline
1 & 6,400 & 0.084\\
4 & 3,200 & 0.136\\
16 & 1,600 & 0.152\\
64 & 800 & 0.152\\
256 & 400 & 0.153\\
1024 & 200 & 0.153\\
\tableline
\end{tabular}
\hspace{1cm}
\begin{tabular}{ccc}
\multicolumn{3}{c}{Run: TCF $\simeq0.60$}\\
\tableline\tableline
$n_{\rm cl}$ & $r_{\rm cl}$ & TCF\\
 & \kms &\\
\tableline
1 & 12,800 & 0.335\\
4 & 6,400 & 0.543\\
16 & 3,200 & 0.607\\
64 & 1,600 & 0.609\\
256 & 800 & 0.610\\
1024 & 400 & 0.607\\
\tableline
\end{tabular}
\hspace{1cm}
\begin{tabular}{ccc}
\multicolumn{3}{c}{Run: TCF $\simeq2.4$}\\
\tableline\tableline
$n_{\rm cl}$ & $r_{\rm cl}$ & TCF\\
 & \kms &\\
\tableline
1 & 24,600 & 2.11\\
4 & 12,800 & 2.35\\
16 & 6,400 & 2.42\\
64 & 3,200 & 2.43\\
256 & 1,600 & 2.44\\
1024 & 800 & 2.43\\
\tableline
\end{tabular}

\end{center}
\end{table}


\end{document}